\documentclass[usenatbib]{mn2e}

\usepackage{aas_macros}
\usepackage{amsmath}
\usepackage{amssymb}
\usepackage[colorlinks=true,breaklinks=true,citecolor=blue]{hyperref}
\usepackage{xspace}
\usepackage{color}
\usepackage{graphicx}
\usepackage{txfonts} 

\newcommand{\eg}{\mbox{\textit{e.\,g.},}\xspace} 
\newcommand{\ie}{\mbox{\textit{i.\,e.},}\xspace}

\newcommand{\COBE}{\textit{COBE}}
\newcommand{\WMAP}{\textit{WMAP}}
\newcommand{\Planck}{\textit{Planck}}

\renewcommand{\vec}[1]{\bmath{#1}} 
\newcommand{\mat}[1]{\mathbfss{#1}}

\definecolor{blue}{rgb}{0,0,1} 
\definecolor{yellow}{rgb}{1,0.84,0}
\definecolor{red}{rgb}{1,0,0}
\definecolor{mypink1}{rgb}{0.858, 0.188, 0.478}
\definecolor{darkgreen}{rgb}{0,0.5,0}
\definecolor{orange}{rgb}{1.0,0.4,0}

\title[SILC: CMB temperature map by directional wavelets]{SILC: a new \textit{Planck} Internal Linear Combination CMB temperature map using directional wavelets}
\author[K.~K.~Rogers et al.]
{Keir K.~Rogers,$^{1}$\thanks{E-mail: keir.rogers.14@ucl.ac.uk} Hiranya V.~Peiris,$^{1}$ Boris Leistedt,$^{2,1}$ Jason D.~McEwen$^{3}$
\newauthor and Andrew Pontzen$^{1}$
\\
$^1$Department of Physics \& Astronomy, University College London, Gower Street, London WC1E 6BT, UK\\
$^2$Department of Physics, New York University, 4 Washington Place, New York, NY 10003, USA\\
$^3$Mullard Space Science Laboratory, University College London, Surrey RH5 6NT, UK}

\begin{document}
\maketitle
\begin{abstract}
We present new clean maps of the CMB temperature anisotropies (as measured by {\Planck}) constructed with a novel internal linear combination (ILC) algorithm using directional, scale-discretised wavelets --- Scale-discretised, directional wavelet ILC or SILC. Directional wavelets, when convolved with signals on the sphere, can separate the anisotropic filamentary structures which are characteristic of both the CMB and foregrounds. Extending previous component separation methods, which use the frequency, spatial and harmonic signatures of foregrounds to separate them from the cosmological background signal, SILC can additionally use morphological information in the foregrounds and CMB to better localise the cleaning algorithm. We test the method on {\Planck} data and simulations, demonstrating consistency with existing component separation algorithms, and discuss how to optimise the use of morphological information by varying the number of directional wavelets as a function of spatial scale. We find that combining the use of directional and axisymmetric wavelets depending on scale could yield higher quality CMB temperature maps. Our results set the stage for the application of SILC to polarisation anisotropies through an extension to spin wavelets.

\end{abstract}

\section{Introduction}
\label{sec:intro}

Accurate measurements of the cosmic microwave background (CMB) arguably form the bedrock of modern precision cosmology. In particular, the full-sky multifrequency CMB maps provided by three generations of satellite experiments  --- {\COBE} \citep{1990ApJ...354L..37M,1992ApJ...397..420B}, {\WMAP} \citep{2003ApJS..148....1B} and {\Planck} \citep{2011A&A...536A...1P} --- represent milestones in our understanding of the cosmological model. However, to obtain a full-sky map of the CMB requires removing instrumental noise and signals due to astrophysical foregrounds (primarily in the Milky Way). Full-sky foreground-cleaned CMB maps are used for a wide variety of scientific purposes \citep[see \eg][]{2015arXiv150607135P,2015arXiv150201592P}.

There are numerous methods to perform foreground component separation. They broadly fall into two categories: blind methods which make minimal physical assumptions about the contributing signals and the so-called mixing matrix (which quantifies the strength of different components at different frequencies) and non-blind methods which are based on a physical modelling of the sky components. Examples of non-blind methods include the Maximum Entropy Method (MEM) \citep{1998MNRAS.300....1H} and the parametric Bayesian CMB Gibbs sampler Commander \citep{2006ApJ...641..665E,2008ApJ...676...10E}. Correlated Component Analysis (CCA) \citep{2005EJASP2005.2400B} is a semi-blind method that estimates the mixing matrix based on second-order statistics. Spectral Estimation via Expectation Maximisation (SEVEM) \citep{2003MNRAS.345.1101M,2008A&A...491..597L,2012MNRAS.420.2162F} is a template fitting technique. Examples of so-called blind source separation include the sparsity-based method Local-Generalized Morphological Component Analysis (L-GMCA) \citep{2008StMet...5..307B,2013A&A...550A..73B} and the Spectral Matching Independent Component Analysis (SMICA) \citep{2008arXiv0803.1814C}, although the latter work does discuss how the choice of component model affects the blindness of this method. Of particular interest to this work is another blind method, the Internal Linear Combination (ILC), most recently implemented by the Needlet ILC (NILC) \citep{2009A&A...493..835D}. In its component separation analysis, the {\Planck} Collaboration used Commander, NILC, SEVEM and SMICA \citep{2015arXiv150205956P}. See, \eg \citet{2009A&A...493..835D,2013A&A...550A..73B} for reviews of CMB component separation methods.

The ILC computes a weighted sum of CMB maps as measured at multiple frequencies. These weights are constrained to sum to unity at each point in the map, ensuring that the CMB signal is conserved, assuming that it is equal at each frequency. Under this constraint, the weights are calculated by minimising the empirical variance of the ILC map, which in turn minimises the variance of the error in CMB reconstruction (assuming the CMB and foregrounds and the CMB and noise are respectively uncorrelated). The variances we minimise are empirical in that they are calculated using the data themselves. In this way, the weights are calculated to remove foreground and noise, revealing the underlying primordial CMB anisotropies. The ILC method was originally used by the {\WMAP} Collaboration \citep{2003ApJS..148...97B} and then extended by \citet{2004ApJ...612..633E} through an analytical calculation of the weights. One limitation of the original ILC approach is the extent of localisation of the weights. The initial versions calculated different weights in separate parts of the sky (\eg \citealt{2003ApJS..148...97B} split the Galactic region into 11 parts). In order to further remove local contamination, the weights can be allowed to vary across the sky and also at different spatial scales. \citet{2003PhRvD..68l3523T} made an ILC map allowing the weights to vary at each multipole, as well as within different regions of the sky. A direct extension of this work is to make use of both spatial and frequency information simultaneously using wavelets. The weights are then defined across wavelet scales and within wavelet coefficient maps on the sky.

Wavelets are functions that are localised in both real and frequency space. To analyse full-sky CMB maps, wavelets defined on the surface of a sphere are required. A number of wavelet frameworks on the sphere have been developed recently \citep{antoine:1999, antoine:1998, baldi:2009, barreiro:2000,geller:2008, geller:2010:sw, geller:2010, 2013A&A...558A.128L, 2015arXiv150203120L, 2006SIAM...38...574, mcewen:2006:cswt2, mcewen:2008:fsi, mcewen:szip, 2013SPIE.8858E..0IM, 2014IAUS..306...64M, mcewen:s2let_localisation, mcewen:s2let_spin, marinucci:2008, sanz:2006, starck:2006, starck:2009,wiaux:2005, 2008MNRAS.388..770W}. In particular, needlets \citep{2006SIAM...38...574,marinucci:2008,baldi:2009} have been used in the latest generation of ILC methods. Needlets are a set of axisymmetric kernels defined on the surface of a sphere. Each member of the set has compact support in harmonic space over different multipole ranges.  When each needlet is convolved with a signal defined on the sphere, the resulting signal (\ie needlet coefficients) also has compact harmonic support. NILC \citep{2009A&A...493..835D} computes its weights by considering needlet scales separately (harmonic localisation) and then different parts of each needlet coefficient map separately (spatial localisation). The needlets are constructed in such a way that the original signal can be recovered from its needlet coefficients with no loss of information (in practice, small losses can be introduced by approximate spherical harmonic transforms). NILC has been very successful at forming clean full-sky CMB maps, which contain very little residual foreground and noise contamination.

In this work, we introduce the Scale-discretised, directional wavelet ILC or SILC, which extends the wavelet ILC framework by localising the calculation of ILC weights in an additional domain. We use wavelets which are not only harmonically-localised but also directional \citep{2008MNRAS.388..770W,2013SPIE.8858E..0IM,mcewen:s2let_spin}. Unlike needlets, which are axisymmetric on the sphere, directional wavelets are non-axisymmetric, \ie the kernels are ``squeezed.'' This means that for one wavelet scale, one axisymmetric kernel is replaced by a number of complementary directional kernels, each with a different orientation. When these directional wavelets are convolved with a signal on the sphere, (within each scale) different orientations of signal structure are separated. This directional localisation allows the ILC weights to be additionally fine-tuned to better remove foreground and noise, in particular for signals with filamentary structure. Furthermore, directional wavelets exhibit exact reconstruction, allowing them to be embedded in an ILC such that no signal is lost.

SILC is being developed with the goal of analysing CMB polarisation components through an extension to spin, directional wavelets \citep{mcewen:s2let_spin, 2014IAUS..306...64M,2015arXiv150203120L}, which are expected to be well-suited to localising the complex filamentary morphologies of polarised foregrounds. As a precursor step, in this work we test SILC on the scalar temperature field in order to demonstrate the quantitative consistency of its foreground cleaning performance compared with existing component separation methods, and to identify possible optimisations for the extension to spin fields.

Directional wavelets are explained briefly in \S~\ref{sec:dir_wav}. In \S~\ref{sec:method}, the SILC algorithm is explained in detail. Various sources of error in the method are considered in \S~\ref{sec:ilc_error}. In \S~\ref{sec:comp}, we compare our method to previous component-separation methods. The application to {\Planck} simulations (\S~\ref{sec:simulations}) is followed by application to {\Planck} data (\S~\ref{sec:data}). We discuss the results and error estimation based on the data  in \S~\ref{sec:discussion} and conclude in \S~\ref{sec:concs}.

\section{Directional wavelets}
\label{sec:dir_wav}

\begin{figure}
\includegraphics[width=\columnwidth]{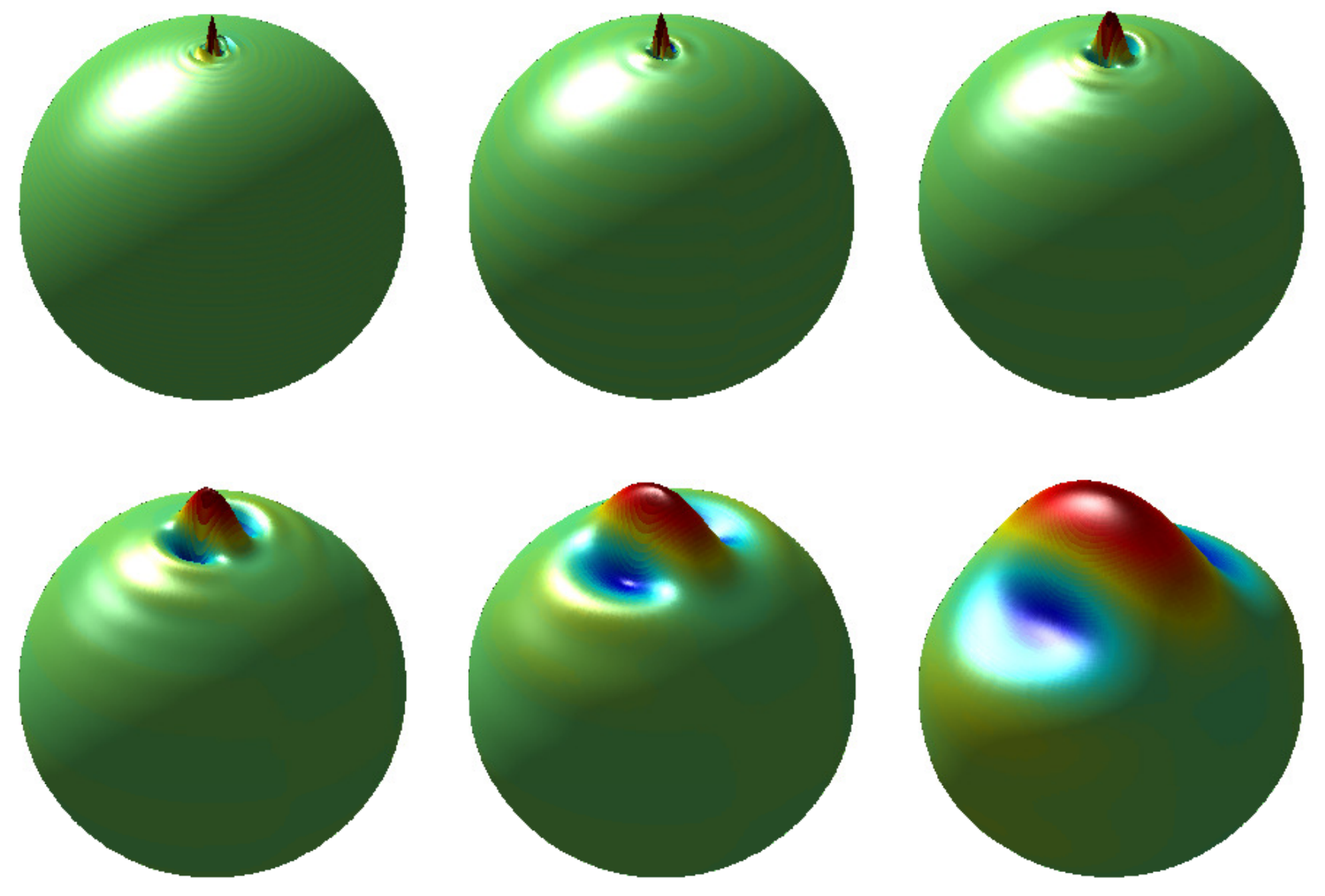}
\caption{The spatial localisation on the sphere of directional, scale-discretised wavelets. Each sub-plot shows a representation of a directional wavelet kernel at different scales, where red, raised parts show positive wavelet response and blue, depressed parts show negative wavelet response. \textit{From left to right, top to bottom}: wavelet scale index \(j\) decreases. The number of directions per wavelet scale \(N = 3\). Therefore, for complete reconstruction at each scale, the above wavelets would be complemented by two more wavelets of the same size but of a different orientation on the sphere. This figure is adapted from \citet{2013SPIE.8858E..0IM}.}
\label{fig:wavelets_spatial}
\end{figure}

\begin{figure}
\includegraphics[width=\columnwidth]{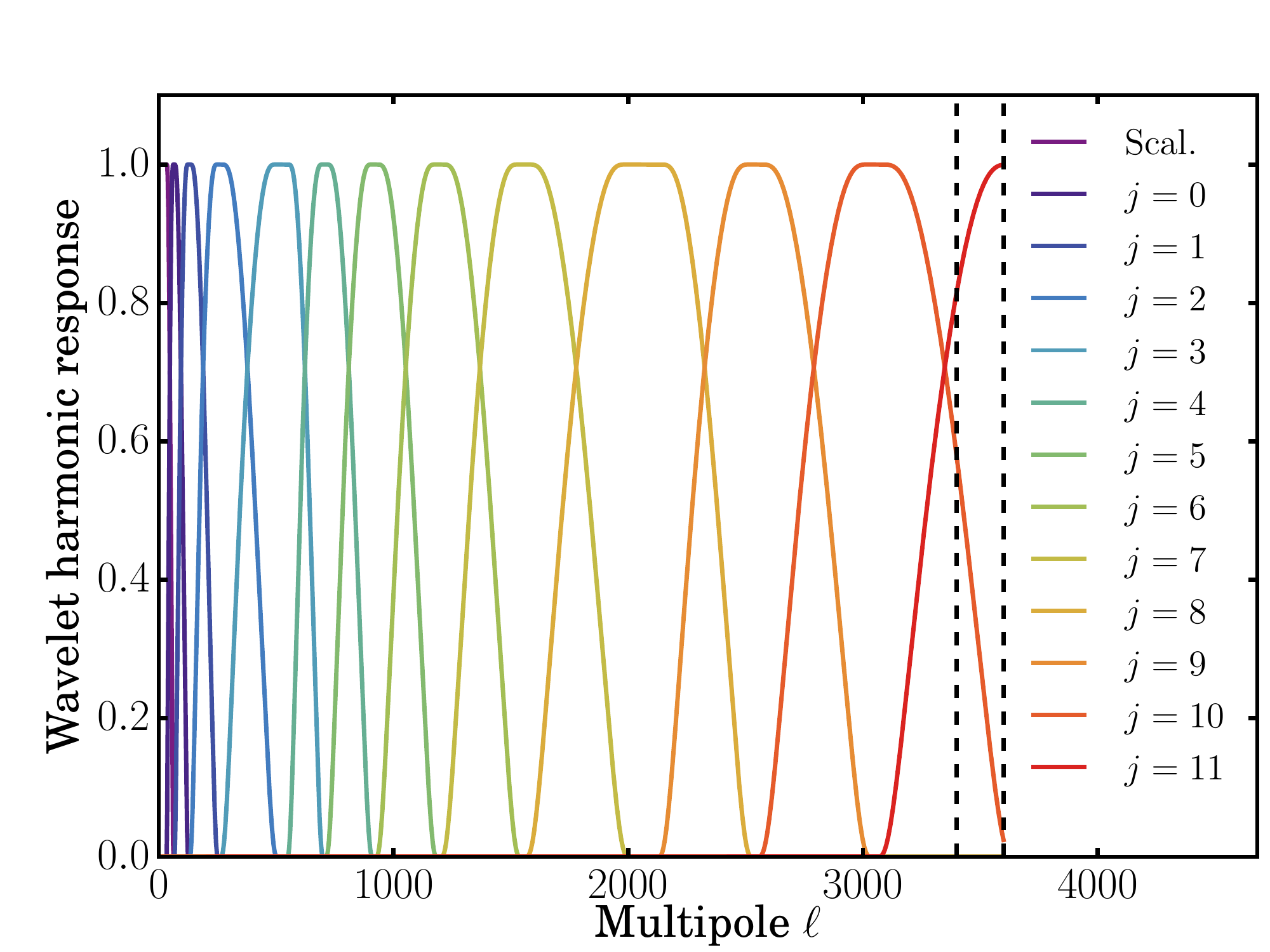}
\caption{The harmonic response of the directional wavelets used in this work, where \(j\) specifies the wavelet scale.  Increasing \(j\) corresponds to a smaller wavelet kernel and so a multipole range on smaller scales (\ie larger multipoles $\ell$). The largest wavelet scale (\(\mathrm{Scal.}\)) is the scaling function (\S~\ref{sec:wavelets}). The two smallest wavelets are harmonically truncated at \(\ell = 3600\) but are smoothly tapered to zero from \(\ell = 3400\) to \(\ell = 3600\) (the two dotted lines) by the beam tapering discussed in \S~\ref{sec:beam}. The band-limits of the above wavelets are given in Table \ref{tab:wav_lims}.}
\label{fig:wavelets}
\end{figure}

\begin{table}\centering
\caption{The harmonic band-limits \([\ell^j_{\mathrm{min}},\ell^j_{\mathrm{max}}]\) of the directional wavelets used in this work. \(\ell^j_{\mathrm{peak}}\) is the multipole at which each wavelet has its maximum response. The final column shows the number of equiangular samples per wavelet coefficient map \(N^j_\mathrm{samp}\).}
\label{tab:wav_lims}
\begin{tabular}{ccccc}
\hline
Wavelet scale \(j\) & \(\ell^j_{\mathrm{min}}\) & \(\ell^j_{\mathrm{peak}}\) & \(\ell^j_{\mathrm{max}}\) & \(N^j_\mathrm{samp}\) \\
\hline
Scal. & 0 & 64 & 64 & 8,385 \\
0 & 32 & 64 & 128 & 33,153 \\
1 & 64 & 128 & 256 & 131,841 \\
2 & 128 & 256 & 512 & 525,825 \\
3 & 256 & 512 & 706 & 998,991 \\
4 & 542 & 705 & 918 & 1,688,203 \\
5 & 705 & 917 & 1193 & 2,850,078 \\
6 & 917 & 1192 & 1551 & 4,815,856 \\
7 & 1192 & 1550 & 2015 & 8,126,496 \\
8 & 1550 & 2015 & 2540 & 12,910,821 \\
9 & 2116 & 2539 & 3048 & 18,589,753 \\
10 & 2539 & 3047 & 3600 & 25,930,801 \\
11 & 3047 & 3600 & 3600 & 25,930,801 \\
\hline
\end{tabular}
\end{table}

\begin{figure}
\begin{tabular}{c}
\includegraphics[width=\columnwidth]{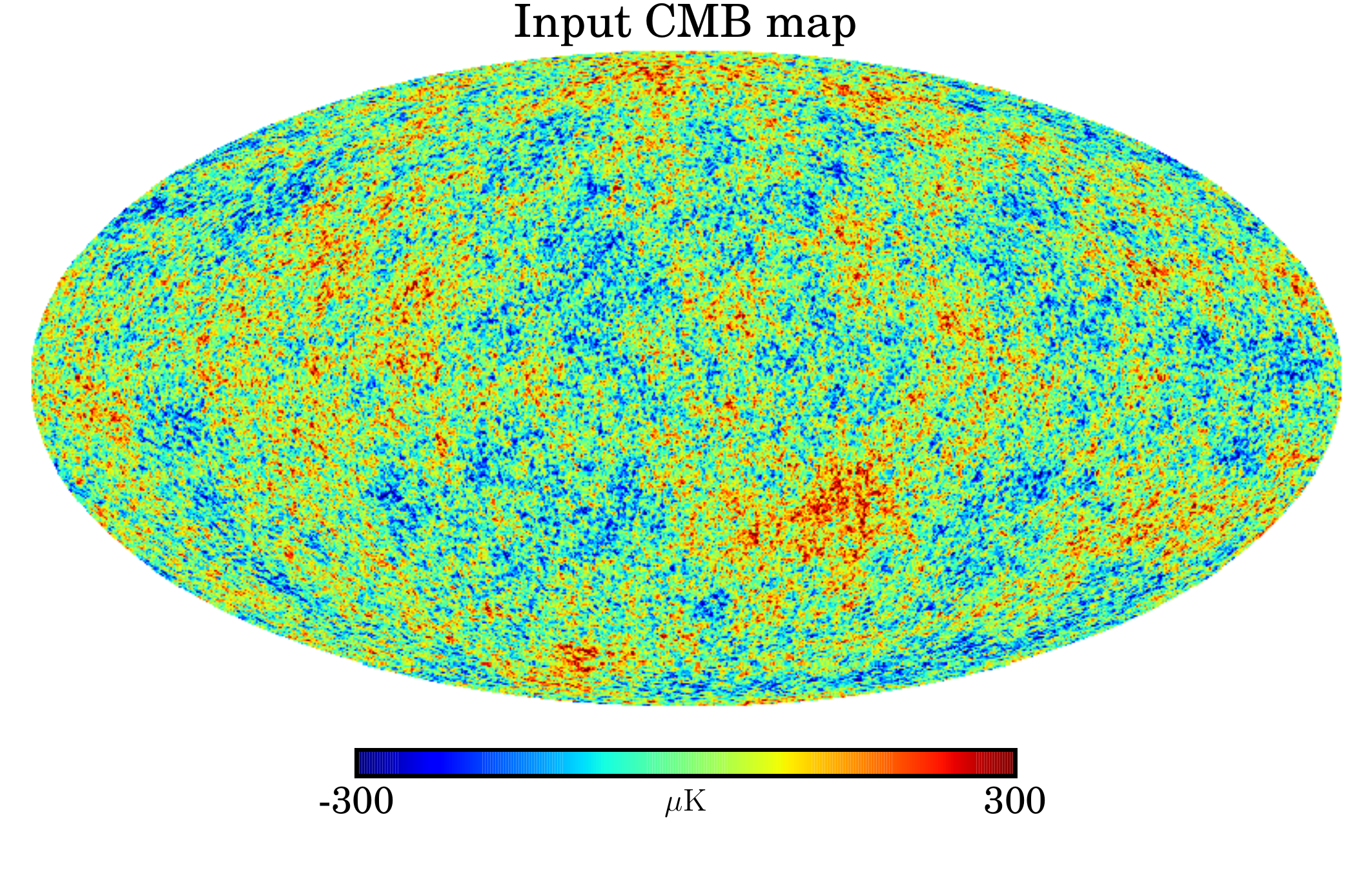} \\
\includegraphics[width=\columnwidth]{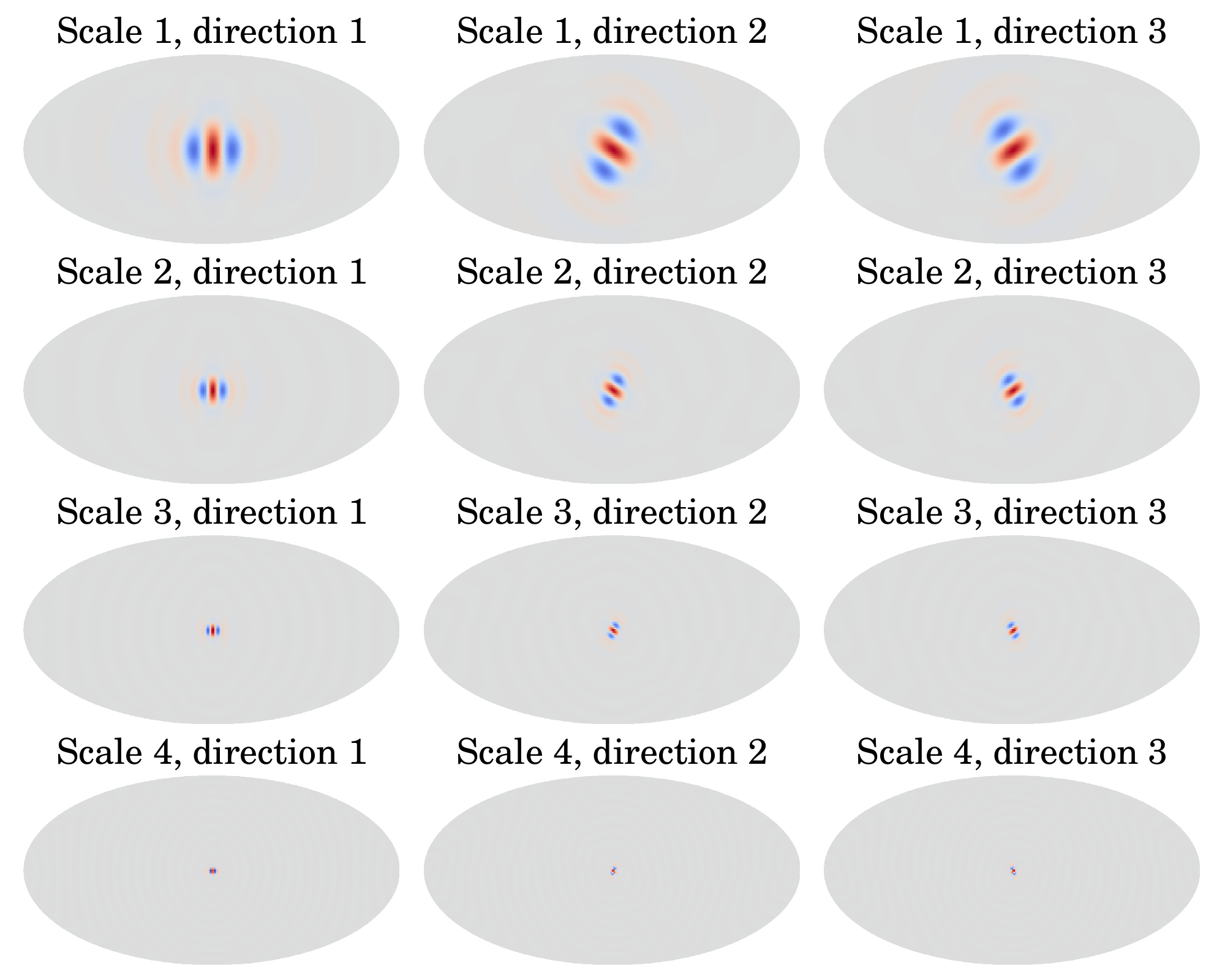} \\
\includegraphics[width=\columnwidth]{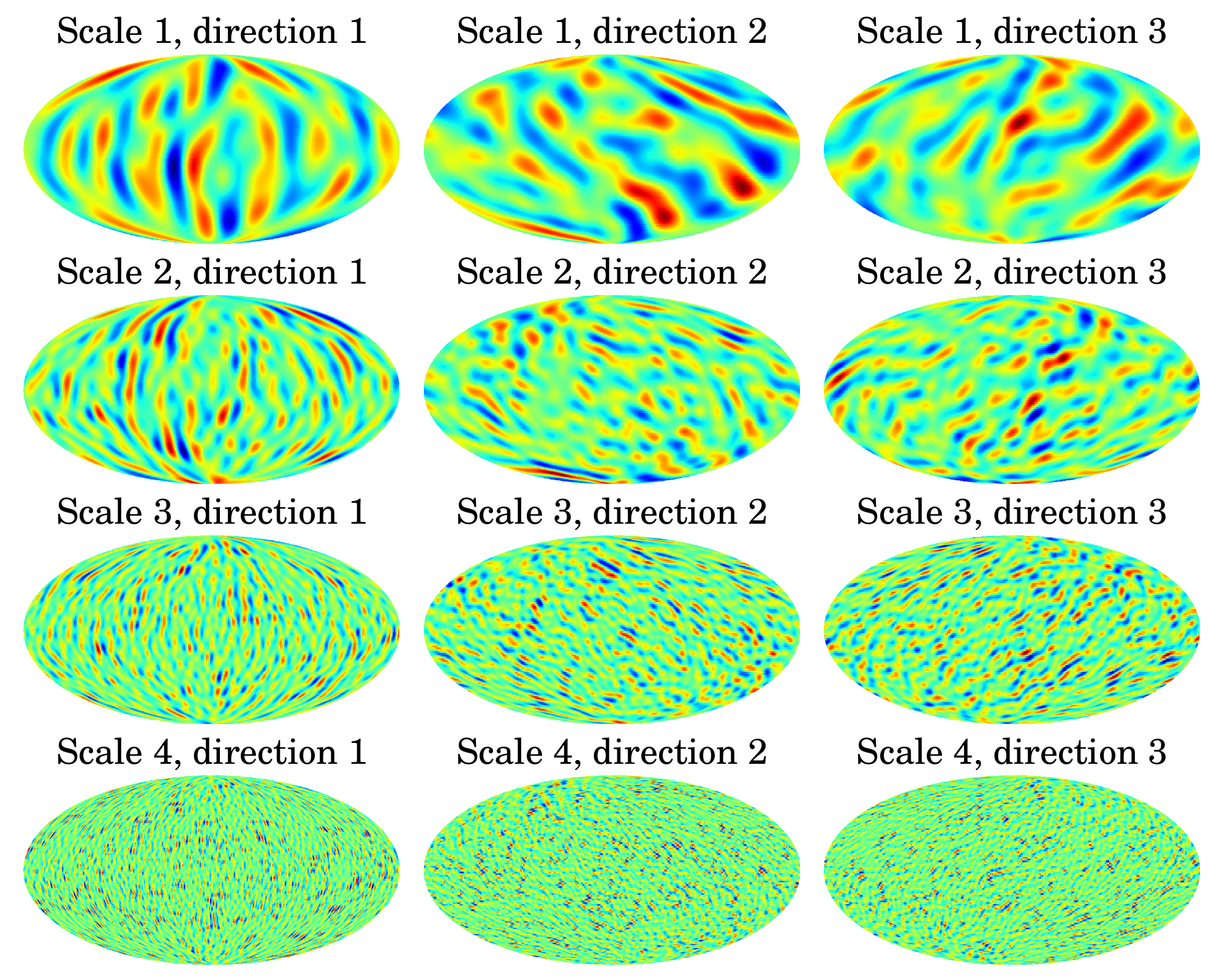}
\end{tabular}
\caption{The CMB (\textit{top map}) decomposed into directional wavelet coefficient maps (\textit{bottom section}). The wavelet kernels are shown (\textit{middle section}), where red indicates positive response and blue indicates negative. In the full analysis, we also include smaller wavelets than we show above.}
\label{fig:cmb_decomp}
\end{figure}

Directional, scale-discretised wavelets on the sphere that support exact reconstruction have been developed in \citet{2008MNRAS.388..770W}, \citet{2013SPIE.8858E..0IM} and \citet{mcewen:s2let_spin}, while their localisation properties have been studied in \citet{mcewen:s2let_localisation}. Figure \ref{fig:wavelets_spatial} shows an example of the spatial localisation of directional wavelets. Larger wavelet scales have larger kernels, and when these are convolved with signals defined on the sphere (such the CMB and astrophysical foregrounds), signal structure with the same scale and orientation as the wavelet is isolated. The kernels in Fig.~\ref{fig:wavelets_spatial} are shown for a single direction and (for complete reconstruction) would be complemented by two more sets of kernels of the same sizes but rotated to different orientations. Figure \ref{fig:wavelets} shows an example of the harmonic localisation of directional wavelets (for the wavelets used in this work). The harmonic supports of the wavelets overlap, with each wavelet covering a finite set of multipoles. Figure \ref{fig:cmb_decomp} shows an example of directional wavelet decomposition as applied to the CMB. Although the CMB anisotropies are statistically Gaussian, the CMB spots on the sky demonstrate anisotropy as a function of scale \citep{1987MNRAS.226..655B}. When the CMB is convolved with directional wavelets, structure of different orientations is separated. This further supports the use of directional wavelets in CMB analysis: both filamentary foreground structure and the CMB itself are better localised. This particularly applies in the case of polarisation, as will be discussed in \S~\ref{sec:concs} when we consider extensions to our method. For a mathematical description of directional wavelets, see \S~\ref{sec:wavelets}.

In the spherical harmonic transforms used in the computation of directional wavelet coefficient maps, we adopt the sampling scheme on the sphere of \citet{2011ITSP...59.5876M} (hereafter MW sampling), rather than, \eg HEALPix sampling \citep{2005ApJ...622..759G}, although in principle HEALPix could be used if desired. The corresponding sampling theorem of \citet{2011ITSP...59.5876M} shows that the MW sampling scheme requires fewer samples for a band-limited signal than any other sampling theorem. Additionally, the use of a separation of variables and fast Fourier transforms (FFTs) yields a numerically efficient algorithm. In particular, our spherical harmonic transforms are theoretically exact, unlike HEALPix. This allows one to manipulate signals with the minimal number of samples and to perform the numerous spherical harmonic transforms involved in the ILC algorithm without any loss of information (other than that due to the finite representation of floating point numbers). Our final map products, however, are provided in HEALPix format. Finally, MW sampling of spin signals on the sphere requires no additional computational complexity and this will be vital in the extension of our method to polarisation \(E\) and \(B\) modes (\S~\ref{sec:concs}). Further details on MW sampling are given in \S~\ref{sec:wavelets}.

\section{Method}
\label{sec:method}

We start by outlining the SILC algorithm. The steps are explained in more detail in the subsequent subsections (\S~\ref{sec:input} to \ref{sec:ps}). We discuss our numerical implementation in \S~\ref{sec:num_imp}.
\begin{enumerate}
\renewcommand{\theenumi}{(\arabic{enumi})}
\item The raw input data are multifrequency full-sky maps of CMB temperature fluctuations. These maps use the HEALPix format. (See \S~\ref{sec:input}.) The model we employ for the raw data is explained in \S~\ref{sec:model}.
\item The maps are ``pre-processed'' by inpainting in a small point source mask (see \S~\ref{sec:ps}).
\item The input maps are converted to thermodynamic (CMB) temperature (if necessary). For {\Planck} temperature data, the 545 GHz and 857 GHz maps are converted from spectral flux density per unit solid angle (MJy Sr\(^{-1}\)) to CMB temperature (K\(_{\mathrm{CMB}}\)) by the unit conversions given in the {\Planck} 2015 Release Explanatory Supplement\footnote{{\Planck} 2015 Release Explanatory Supplement: UC CC tables (\url{http://wiki.cosmos.esa.int/planckpla2015/index.php/UC_CC_Tables}). For the 545 GHz map, the unit conversion is (58.0356 \(\pm\) 0.0278) MJy Sr\(^{-1}\) K\(_{\mathrm{CMB}}^{-1}\) and for the 857 GHz map, the unit conversion is (2.2681 \(\pm\) 0.0270) MJy Sr\(^{-1}\) K\(_{\mathrm{CMB}}^{-1}\).}.
\item The maps are each convolved to have the same effective beam (see \S~\ref{sec:beam}).
\item Each input map is converted into a set of wavelet coefficient maps. This separates both the scale and orientation of structure within each map. These wavelet coefficient maps use MW sampling \citep{2011ITSP...59.5876M}. (See \S~\ref{sec:wavelets}.)
\item The ILC method is then applied separately to each wavelet scale and orientation. For each scale and orientation, the multifrequency wavelet coefficient maps are weighted and added to form a single wavelet coefficient map that contains mainly CMB signal, as well as some residual foreground and noise. These weights are allowed to vary from wavelet coefficient to wavelet coefficient. The calculation of these weights is explained in \S~\ref{sec:ilc}.
\item The final ILC wavelet coefficient maps are synthesised to form the final product: a full-sky map of CMB temperature fluctuations (with some residual foreground and noise). The final map uses the HEALPix format. (See \S~\ref{sec:wavelets}.)
\item The final map is inpainted in a small point source mask (see \S~\ref{sec:ps}).
\end{enumerate}

\subsection{Input data}
\label{sec:input}

Our main CMB temperature map products use full-mission 2015 release {\Planck} temperature maps as their input\footnote{\url{http://pla.esac.esa.int/pla}}. All nine frequency channels are used. At 70 GHz, we use the higher-resolution version at \(N_{\mathrm{side}} = 2048\). We also use the full-mission Full Focal Plane 8 (FFP8) simulations \citep{2015arXiv150906348P} without bandpass mismatch\footnote{FFP8 simulations are also available with bandpass mismatch, accounting for differences in the bandpasses of detectors nominally at the same frequency, leading to spurious signals in the frequency maps.}.

\subsection{Data model}
\label{sec:model}

Each full-sky temperature map can be modelled \citep[\eg][]{2012MNRAS.419.1163B} as
\begin{equation}\label{eq:model}
T^{\mathrm{OBS},c}(\hat{\vec{n}}) = \int_{\hat{\vec{n}}'} \mathrm{d}\hat{\vec{n}}' B^c(\hat{\vec{n}},\hat{\vec{n}}') T^{\mathrm{SIG},c}(\hat{\vec{n}}') + T^{\mathrm{N},c}(\hat{\vec{n}}),
\end{equation}
where the signal component can further be decomposed as
\begin{equation}\label{eq:signal}
T^{\mathrm{SIG},c}(\hat{\vec{n}}) = a^c T^{\mathrm{CMB}}(\hat{\vec{n}}) + T^{\mathrm{FG},c}(\hat{\vec{n}}).
\end{equation}
\(T^{\mathrm{CMB}}(\hat{\vec{n}})\) is the CMB component at a point on the sky \(\hat{\vec{n}}\). \(T^{\mathrm{FG},c}(\hat{\vec{n}})\) and \(T^{\mathrm{N},c}(\hat{\vec{n}})\) are respectively the foreground and detector noise components for frequency channel \(c\). \(a^c\) is the calibration coefficient for the CMB for each channel. The overall signal component is smoothed by a beam function \(B^c(\hat{\vec{n}},\hat{\vec{n}}')\) due to the finite resolution of the observations. However, the noise component is not smoothed by the beam. Here we assume the beam to be circularly symmetric. Therefore, the beam can be represented as a sum over Legendre polynomials,
\begin{equation}\label{eq:beam}
B^c(\hat{\vec{n}},\hat{\vec{n}}') = \sum_{\ell=0}^\infty \frac{2\ell+1}{4\pi} B_\ell^c P_\ell(\hat{\vec{n}}.\hat{\vec{n}}').
\end{equation}
We can recast Eq.~\eqref{eq:model} in the spherical harmonic representation as
\begin{equation}\label{eq:model_alm}
a_{\ell m}^{\mathrm{OBS},c} = a^c B_\ell^c a_{\ell m}^\mathrm{CMB} + B_\ell^c a_{\ell m}^{\mathrm{FG},c} + a_{\ell m}^{\mathrm{N},c}
\end{equation}
where \(a_{\ell m}\) are the coefficients of spherical harmonics \(Y_{\ell m}(\hat{\vec{n}})\).

\subsection{Beam convolution}
\label{sec:beam}

Equation \eqref{eq:model_alm} shows that each frequency channel \(c\) has a different beam transfer function \(B_\ell^c\). To replace each beam with a channel-independent resolution, we perform a deconvolution/convolution procedure to give spherical harmonic coefficients
\begin{equation}\label{eq:conv}
a_{\ell m}^c = \frac{B_\ell^\mathrm{EFF}}{B_\ell^c} a_{\ell m}^{\mathrm{OBS},c} \, ,
\end{equation}
where \(B_\ell^\mathrm{EFF}\) is the final (effective) beam transfer function of our map products. For {\Planck} data, we use a Gaussian beam with a FWHM of 5\arcmin\ as our input beam. We taper this beam to zero from \(\ell = 3400\) to \(\ell = 3600\) using a Fermi function. This suppresses any small-scale power aliasing due to having harmonically-truncated wavelets in this multipole range. Convolving with beam transfer functions ignores the non-axisymmetric component of the beams; these will remain in the input maps but are assumed to be small.

This deconvolution/convolution procedure does not correctly handle the noise component of our input maps. Equation \eqref{eq:conv} can be expanded (using Eq.~\eqref{eq:model_alm}) as
\begin{equation}\label{eq:conv_expand}
a_{\ell m}^c = B_\ell^\mathrm{EFF} (a^c a_{\ell m}^\mathrm{CMB} + a_{\ell m}^{\mathrm{FG},c}) + \frac{B_\ell^\mathrm{EFF}}{B_\ell^c} a_{\ell m}^{\mathrm{N},c},
\end{equation}
where \(a^c\) are the CMB calibration coefficients (not to be confused with the inverse spherical harmonic transform of harmonic coefficients \(a_{lm}^c\)). The final resolution of an ILC map is usually chosen to match the best resolution of the input maps. Therefore, for all but the highest resolution channel and for all \(\ell\), \(B_\ell^\mathrm{EFF} > B_\ell^c\). This has the effect of increasing the noise contribution of the input maps, particularly at high \(\ell\) and for low-resolution maps, where \(B_\ell^\mathrm{EFF} \gg B_\ell^c\). We use the {\Planck} beam transfer functions as provided in the Reduced Instrument Model (RIMO)\footnote{{\Planck} 2015 Release Explanatory Supplement: The 2015 instrument model (\url{http://wiki.cosmos.esa.int/planckpla2015/index.php/The_RIMO}).}. For the LFI beams, we use Gaussian approximations with FWHM {32.33\arcmin}, {27.01\arcmin} and {13.25\arcmin} for 30, 44 and 70 GHz respectively. Following \citet{2014A&A...571A..12P}, the deconvolved beams are thresholded such that the \(B_l^c\) is set to the value given in the RIMO or 0.001, whichever is larger. This prevents the last term in Eq.~\eqref{eq:conv_expand} from becoming so large that numerical errors occur. Although we lose accuracy in the deconvolution process, the contribution of the channels in the multipole ranges affected is highly attenuated in the ILC weights in any case.

\subsection{Wavelet analysis and synthesis}
\label{sec:wavelets}

The wavelet ILC method requires the decomposition of each band-limited temperature map \(T^c(\hat{\vec{n}})\) into a set of wavelet coefficient maps \(W^{\Psi^j}\): in our case, directional, scale-discretised wavelets \citep{2008MNRAS.388..770W,2013SPIE.8858E..0IM,mcewen:s2let_spin}. A general introduction was provided in \S~\ref{sec:dir_wav} --- here we provide some technical details of the implementation. We drop the \(c\) superscript on \(T\) for the rest of this subsection since each map is analysed using the same wavelets.  The wavelet coefficients are defined as the directional convolution of $T$ with wavelets defined on the sphere \(\Psi^j \in \mathrm{L}^2(\mathbb{S}^2)\) (specifically those shown in Fig.~\ref{fig:wavelets}) where index $j$ denotes the wavelet scale. Importantly, directional wavelets yield coefficients \( W^{\Psi^j} (\hat{\vec{\rho}}) \) that live on the space of three-dimensional rotations, \ie the rotation group SO(3):
\begin{equation}\label{eq:wav_coeffs}
W^{\Psi^j} (\hat{\vec{\rho}}) \equiv \langle T, \mathcal{R}_{\hat{\vec{\rho}}} \Psi^j \rangle = \int_{\mathbb{S}^2} \mathrm{d}\hat{\vec{n}} \ T(\hat{\vec{n}}) (\mathcal{R}_{\hat{\vec{\rho}}} \Psi^j)^{*} (\hat{\vec{n}}),
\end{equation}
where $\mathrm{d}\hat{\vec{n}}$ is the usual invariant measure on the sphere, $\cdot^\ast$ denotes complex conjugation and the rotation operator is defined by
\begin{equation}\label{eq:rot}
(\mathcal{R}_{\hat{\vec{\rho}}} \Psi^j)(\hat{\vec{n}}) \equiv \Psi^j (\mat{R}_{\hat{\vec{\rho}}}^{-1} \hat{\vec{n}}),
\end{equation}
where \(\mat{R}_{\hat{\vec{\rho}}}\) is the three-dimensional rotation matrix corresponding to \(\mathcal{R}_{\hat{\vec{\rho}}}\). In these equations, \(\hat{\vec{\rho}} = (\theta,\phi,\chi) \in \mathrm{SO}(3)\) denotes the Euler angles (in the \(zyz\) convention) with colatitude \(\theta \in [0,\pi]\), longitude \(\phi \in [0,2\pi)\) and direction \(\chi \in [0,2\pi)\)\footnote{We adopt the $zyz$ Euler convention corresponding to the rotation of a physical body in a \emph{fixed} coordinate system about the $z$, $y$ and $z$ axes by $\chi$, $\theta$ and $\phi$, respectively.}. In other words, the wavelet coefficients probe directional structure in \(T\) with \(\chi\) corresponding to the orientation about each point $(\theta,\phi)$ on the sphere.

Following the directional construction of scale-discretised wavelets \citep{2008MNRAS.388..770W,2013SPIE.8858E..0IM,mcewen:s2let_spin}, wavelets are defined by their spherical harmonic coefficients in factorised form:
\begin{equation}\label{eq:wavelets}
\Psi_{\ell n}^j \equiv \kappa^j(\ell) s_{\ell n},
\end{equation}
where \(\kappa^j(\ell)\) sets the harmonic localisation (Fig.~\ref{fig:wavelets}) and \(s_{\ell n}\) sets the directional localisation.

In the original definition of scale-discretised wavelets, the size of all harmonic kernels (setting the harmonic localisation of the wavelets) is parameterised by a unique wavelet dilation parameter \(\lambda \in \mathbb{R}_{*}^{+}\), \(\lambda > 1\). Similarly, the number of directions is set by a unique azimuthal band-limit \(N\). These two parameters respectively characterise $\kappa^j(\ell)$ and $s_{\ell n}$ for all $j$. In this work, we vary \(\lambda\) as a function of multipole in order to allow more flexible harmonic localisation. We achieve this by defining different values of \(\lambda\) in different multipole regions and then stitching together harmonically-truncated wavelets at the region boundaries. We use the values $\lambda=2,1.3,1.2$ with transitions at the multipoles $\ell=512,2015$. If at a transition multipole the harmonic peak of the larger wavelet doesn't equal the peak of the smaller wavelet, then a small amount of unit response is used so that the two wavelets can be continuously combined. Wavelets constructed in this manner satisfy the standard admissibility criterion required for exact reconstruction. The harmonic tiling of the resulting wavelets is shown in Fig.~\ref{fig:wavelets}. The technical details of the construction of each kernel is described in \citet{mcewen:s2let_localisation}. Finally, we use a single parameter $N$ for all scales, \ie each wavelet is divided into the same number of directions. However, a possible extension of this work is to vary $N$ as a function of scale $j$, \eg by using curvelet kernels \citep{chan:s2let_curvelets} or other directional optimisations.

In the case of a single parameter $\lambda$, the limits of the wavelet harmonic window for scale \(j\) are simply \((\ell_\mathrm{min}^j,\ell_\mathrm{max}^j) = (\lambda^{j-1},\lambda^{j+1})\), with their peak response at \(\lambda^j\). In our hybrid scheme, this property remains but $j$ and $\lambda$ must be adjusted in each harmonic region. The full details of our tiling are given in Table~\ref{tab:wav_lims}.
When the limits of the harmonic windows of the maximum wavelet scales extend beyond the overall band-limit \(\ell_\mathrm{max}\), the windows are truncated at \(\ell_\mathrm{max}\). Finally, note that a scaling function $W^{\Phi}$ is needed to capture the very low frequency content of the signal. It is axisymmetric and the corresponding scaling coefficients therefore live on the sphere. Here we do not give the full details of the construction of the scaling function or the factors $\kappa^j(\ell)$ and $s_{\ell n}$ since these can be straightforwardly reproduced by following previous approaches \citep{2008MNRAS.388..770W,2013SPIE.8858E..0IM,mcewen:s2let_spin} and using Table~\ref{tab:wav_lims}.

To apply the ILC algorithm, the above continuous wavelet coefficients must be discretised. Since they live on the rotation group SO(3), we represent them using the sampling scheme of \citet{2015ISPL...22.2425M}, which is a generalisation of the MW sampling scheme \citep{2011ITSP...59.5876M}. Because our wavelets have well-defined band-limits, this approach allows a multiresolution scheme where each scale is pixellated with a minimal number of samples. In practice, the $j$-th wavelet scale has a band-limit $\ell_\mathrm{max}^j$ and is only evaluated at locations $(\theta_t^j, \phi_p^j, \chi_n)$ with $t \in \{0,1,\ldots,\ell_\mathrm{max}^j\}$, $p \in \{0,1,\ldots,2\ell_\mathrm{max}^j\}$ and \(n \in \{0,1,\ldots,N-1\}\). Although wavelet coefficients are evaluated at discrete samples only, for a band-limited signal they capture the total information content of the underlying continuous wavelet coefficient representation, probed up to harmonic band-limit \(\ell_\mathrm{max}^j\) and azimuthal band-limit $N$. This is thanks to the sampling theory on the rotation group ${\rm SO}(3)$ of \citet{2015ISPL...22.2425M}. In the full ensemble of realisations, the ILC (see \S~\ref{sec:ilc} for details) has no sensitivity to the choice of coordinate convention for directions \(\chi\). In a single realisation, there will be some marginal sensitivity to this choice manifesting in the localisation of the empirical covariances we use. However, this effect is sub-dominant to the choice of \(N\), on which we concentrate our analysis.

After the ILC method (see \S~\ref{sec:ilc}) has been applied to the sets of wavelet coefficient maps, there is one final map \(W^{\Psi^j,\mathrm{ILC}}(\hat{\vec{\rho}})\), for each wavelet scale \(j\), living on SO(3) and including the multiple orientations $\chi_0, \dots, \chi_{N-1}$. The additional axisymmetric scaling coefficients $W^{\Phi,\mathrm{ILC}}(\hat{\vec{n}})$ live on the sphere. The final temperature map \(T^\mathrm{ILC}(\hat{\vec{n}})\) is synthesised by
\begin{equation}\label{eq:wav_syn}
\begin{split}
T^\mathrm{ILC}(\hat{\vec{n}}) = &\int_{\mathbb{S}^2} \mathrm{d}\hat{\vec{n}}' W^{\Phi,\mathrm{ILC}}(\hat{\vec{n}}') (\mathcal{R}_{\hat{\vec{n}}'} \Phi) (\hat{\vec{n}}) \\
&+ \sum_{j=j_\mathrm{min}}^{j_\mathrm{max}} \int_{\mathrm{SO}(3)} \mathrm{d} \hat{\vec{\rho}} \ W^{\Psi^j,\mathrm{ILC}}(\hat{\vec{\rho}}) (\mathcal{R}_{\hat{\vec{\rho}}} \Psi^j) (\hat{\vec{n}}),
\end{split}
\end{equation}
where $\mathrm{d}\hat{\vec{\rho}}$ is the usual invariant measure on the rotation group. This final ILC temperature map is pixellated using the HEALPix format from its spherical harmonic coefficients $T^\mathrm{ILC}_{\ell m}$.

The wavelet analysis and synthesis are performed using the latest version of the \texttt{S2LET}\footnote{\url{http://www.s2let.org}} code \citep{2013A&A...558A.128L,2015ISPL...22.2425M}, which in turn relies on the \texttt{SSHT}\footnote{\url{http://www.spinsht.org}} \citep{2011ITSP...59.5876M} and \texttt{SO3}\footnote{\url{http://www.sothree.org}} \citep{2015ISPL...22.2425M} codes to compute spin spherical harmonics and Wigner transforms exactly and efficiently using the MW sampling scheme. Thanks to the sampling theorem, the wavelet coefficients can be transformed using Wigner transforms without any loss of information \citep{2015ISPL...22.2425M}.

\subsection{ILC method}
\label{sec:ilc}

Following the wavelet analysis of the input maps (see \S~\ref{sec:wavelets}), there is a wavelet coefficient map \(W_{jnk}^c\) for each channel \(c\), scale \(j\) and orientation \(n\) with a pixel index \(k\).  Using this more compact notation, we conflate the scaling coefficient map with the wavelet coefficient maps as the ILC method applies in exactly the same fashion. The ILC estimate of the CMB signal at each wavelet scale and orientation is defined as a weighted sum of the wavelet coefficient maps at that scale and orientation
\begin{equation}\label{eq:ilc_sum}
W_{jnk}^\mathrm{ILC} \equiv \sum_{c=1}^{N_c} \omega_{jnk}^c W_{jnk}^c \, ,
\end{equation}
where \(\omega_{jnk}^c\) are the weights (which are allowed to vary across the scale and orientation of the signal as well as pixel space) and \(N_c\) is the number of input channels.

We impose a constraint on the weights (to ensure that the CMB signal is preserved) such that
\begin{equation}\label{eq:constraint}
\sum_{c=1}^{N_c} a^c \omega_{jnk}^c = 1.
\end{equation}
Assuming that the CMB and foregrounds and the CMB and noise are respectively uncorrelated, the variance of the error in the result is minimised when the variance of the ILC map itself is minimised. The resulting weights are given by
\begin{equation}\label{eq:weights}
\omega_{jnk}^c = \frac{\sum_{c'=1}^{N_c} (R_{jnk}^{-1})^{cc'} a^{c'}}{\sum_{c=1}^{N_c} \sum_{c'=1}^{N_c} a^c (R_{jnk}^{-1})^{cc'} a^{c'}},
\end{equation}
where the true covariance matrices at scale \(j\), orientation \(n\) and pixel \(k\), \((R_{jnk})^{cc'} = \langle W_{jnk}^c W_{jnk}^{c'} \rangle\) (where the angled brackets indicate an ensemble average). For a derivation of Eq.~\eqref{eq:weights}, see \citet{2003PhRvD..68l3523T,2004ApJ...612..633E}.

In this work, we estimate covariance matrices empirically by the following procedure (as used in \citealt{2012MNRAS.419.1163B,2015arXiv150205956P}). We start by calculating at each pixel \(k\):
\begin{equation}\label{eq:covar_approx}
(R_{jnk}^\mathrm{approx})^{cc'} = W_{jnk}^c W_{jnk}^{c'}.
\end{equation}
We then smooth each element of the above matrix by a Gaussian beam \(w_j(k,k')\) in pixel space to form the empirical estimates of covariance matrices
\begin{equation}\label{eq:covar_map}
(\hat{R}_{jnk})^{cc'} = \sum_{k'=1}^{N_\mathrm{samp}^j} w_j(k,k') (R_{jnk'}^\mathrm{approx})^{cc'}
\end{equation}
where \(N_\mathrm{samp}^j\) is the total number of pixels in a given map at scale \(j\). For computational efficiency, we perform this smoothing in harmonic space:
\begin{equation}\label{eq:covar_harm}
(\hat{R}_{jnk})^{cc'} = \sum_{\ell=0}^{2\ell_\mathrm{max}^j} \sum_{m=-\ell}^{\ell} w_j^\ell (r_{jn}^{\ell m})^{cc'} Y_k^{\ell m},
\end{equation}
where \((r_{jn}^{\ell m})^{cc'}\) are the harmonic coefficients of the maps formed by the elements of matrices \((R_{jnk}^\mathrm{approx})^{cc'}\), \(w_j^\ell\) is a Gaussian beam transfer function and \(Y_k^{\ell m}\) are the spherical harmonics evaluated at pixel \(k\).

The size of the Gaussian kernel used to smooth the covariance matrices is chosen to be proportional to the size of the wavelet used to form a particular set of wavelet coefficient maps\footnote{\(\mathrm{FWHM}^j = 50 \sqrt{\frac{1200}{N_\mathrm{samp}^j}}\). This value is the same as used in the NILC implementation on {\Planck} data.}. In general the estimation of covariance matrices in ILC methods could be further optimised. It may be preferable to dynamically adapt the smoothing kernel used based on local data. \citet{2009A&A...493..835D} suggested using a larger kernel at high Galactic latitudes where Galactic emission does not vary so much and a smaller kernel towards the Galactic equator where emission is more complex. It could involve masking equatorial regions when estimating the covariance at higher latitudes (somewhat akin to \citet{2014A&A...571A..12P}). It could involve convolving the maps of elements of covariance matrices with the same directional wavelet in order to pick out how the local covariance follows the directional structure of the underlying signal. As mentioned above, in this work, we use a similar method as in previous work for ease of comparison.

It is also worth discussing the upper limit on the summation over \(\ell\) in Eq.~\eqref{eq:covar_harm}. We first note the general rule that for the product of two spherical harmonics \citep{Driscoll1994202}
\begin{equation}\label{eq:y_lm}
Y_{\ell_1,m_1}(\hat{\vec{n}}) Y_{\ell_2,m_2}(\hat{\vec{n}}) = \sum_{L = |\ell_1 - \ell_2|}^{\ell_1 + \ell_2} a_{L,m_1+m_2} Y_{L,m_1+m_2}(\hat{\vec{n}})
\end{equation}
where \(Y_{L,m_1+m_2}(\hat{\vec{n}})\) is defined to be zero if \(|m_1 + m_2| > L\). It follows that for the product of two band-limited maps \(M(\hat{\vec{n}}) = \sum_{\ell=\ell_1}^{\ell_2} \sum_{m=-\ell}^\ell m_{\ell m} Y_{\ell m}(\hat{\vec{n}})\) and \(N(\hat{\vec{n}}) = \sum_{\ell=\ell_3}^{\ell_4} \sum_{m=-\ell}^\ell n_{\ell m} Y_{\ell m}(\hat{\vec{n}})\) (where, without loss of generality, \(\ell_1 \leq \ell_4\)):
\begin{equation}\label{eq:product}
M(\hat{\vec{n}}) N(\hat{\vec{n}}) = \sum_{L=0}^{\ell_2 + \ell_4} \sum_{M=-L}^L p_{LM} Y_{LM}(\hat{\vec{n}})
\end{equation}
for \(\ell_3 < \ell_2\), \ie the limits on \(\ell\) in the two maps overlap (the \(p_{LM}\) are the new harmonic coefficients). (If the limits do not overlap, the lower limit on \(L\) in Eq.~\eqref{eq:product} becomes \(\ell_3 - \ell_2\).) The limits on \(\ell\) in the wavelet coefficient maps \(W_{jnk}^c\) are \((\ell_\mathrm{min}^j,\ell_\mathrm{max}^j)\) (see \S~\ref{sec:wavelets}). Therefore, by Eq.~\eqref{eq:product} and Eq.~\eqref{eq:covar_map}, the limits on \(\ell\) in the covariance matrix element maps \((\hat{R}_{jnk})^{cc'}\) are \((0,2\ell_\mathrm{max}^j)\); hence the limits on \(\ell\) in Eq.~\eqref{eq:covar_harm}.

Having established the main equations governing the ILC method, we now present the main steps in the ILC algorithm that we use:
\begin{enumerate}
\item Form the \((R_{jnk}^\mathrm{approx})^{cc'}\) by Eq.~\eqref{eq:covar_approx}.
\item Smooth the \((R_{jnk}^\mathrm{approx})^{cc'}\) in harmonic space by Eq.~\eqref{eq:covar_harm}.
\item Take the inverse of each covariance matrix at each pixel to form \((\hat{R}_{jnk}^{-1})^{cc'}\).
\item Calculate the ILC weights \(\omega_{jnk}^c\) by Eq.~\eqref{eq:weights}, where we assume that \(a^c = 1\) for all \(c\) and we substitute the empirical estimates for the inverse covariance matrices.
\item Finally, calculate the ILC estimate wavelet coefficient maps \(W_{jnk}^\mathrm{ILC}\) by applying Eq.~\eqref{eq:ilc_sum}.
\end{enumerate}

\subsection{Point source masking}
\label{sec:ps}

The input frequency maps are diffusively inpainted in a small point source mask following the method employed by \citet{2015arXiv150201592P}. This recognises that the ILC fails when the CMB is obscured by bright extragalactic sources or complex emission near the Galactic equator. The inpainting removes these sources and attempts to replace them with an extrapolation of the surrounding signal. The mask supplied is taken from the NILC section of \citet{2014A&A...571A...1P} and is constructed from the Planck Catalogue of Compact Sources (PCCS) \citep{2014A&A...571A..28P,2015arXiv150702058P}\footnote{The details of its construction are given in \citet{2014A&A...571A...1P}. It can be downloaded from \url{http://pla.esac.esa.int/pla} and is labelled \texttt{I\_MASK} in the NILC data products.}. It masks about 2.2\% of the whole sky, predominantly along the Galactic equator towards the Galactic centre.

Because of this inpainting, the final ILC map is inpainted within the point source mask. For the purposes of this inpainting, we have split the mask into two, based on the size of its constituent individual contiguous holes\footnote{Note that many holes can be large and irregularly-shaped due to the overlapping of smaller circular holes.}. For holes consisting of less than or equal to 800 pixels, we inpaint with a constrained Gaussian realisation following the method of \citet{2013A&A...555A..37B}, itself an approximate implementation of the Hoffman-Ribak algorithm \citep{1991ApJ...380L...5H}. For holes consisting of more than 800 pixels (the largest 131 out of 10031), we inpaint with a standard diffusive algorithm (in particular, following the method employed by \citet{2015arXiv150201592P}). The result is that the ILC map is 1.3\% Gaussian inpainted and 0.9\% diffusively inpainted. This follows \citet{2013A&A...555A..37B}, who do not recommend using their Gaussian inpainting for large holes near the Galactic equator.

\subsection{Numerical implementation}
\label{sec:num_imp}

SILC is implemented in \texttt{Python} and is parallelised. At full {\Planck} resolution (\(N_\mathrm{side} = 2048\), \(\ell_\mathrm{max} = 3600\)), when run on a 60-core symmetric multiprocessor (SMP) with 1.5 TB RAM and a 24-core cluster node with 256 GB RAM\footnote{The exact specification for our infrastructure is an Intel Xeon E7-4890 2.8 GHz SMP with 4 \(\times\) 15-core CPUs with 25.6 GB RAM per core, and an Intel Xeon E5-2697 2.7 GHz node with 2 \(\times\) 12-core CPUs with 10.7 GB RAM per core.}, the pipeline takes approximately 12 hours per direction. As shown by Eq.~\eqref{eq:covar_harm}, we perform spherical harmonic transforms to \(2 \ell_\mathrm{max} = 7200\). For a given number of directions \(N\), the full pipeline takes approximately \(N\) times as long as the axisymmetric limit of our method (when \(N = 1\)). In our infrastructure, the code was usually memory-limited (due to the very high resolution of the covariance matrix maps \((R_{jnk})^{cc'}\) at harmonic band-limit \(2 \ell_\mathrm{max}^j\)); the amount of parallelisation sometimes had to be reduced to prevent memory overloads on a single node. As mentioned in \S~\ref{sec:wavelets}, the wavelet transforms employ the latest version of \texttt{S2LET} \citep{2013A&A...558A.128L,2015ISPL...22.2425M}, written in \texttt{C} with \texttt{Python} wrappers, itself employing \texttt{SSHT} \citep{2011ITSP...59.5876M} and \texttt{SO3} \citep{2015ISPL...22.2425M}. Despite the use of MW sampling and FFTs, spherical harmonic transforms are the most time-consuming part of the pipeline, again due to the very high resolution of the \((R_{jnk})^{cc'}\) (for the smallest wavelets, these covariance maps are band-limited at \(\ell = 7200\)). There is scope to further optimise the implementation. Our wavelet analysis and synthesis functions do not respectively output and take as input wavelet coefficient maps at double-resolution (\ie a map band-limited at \(\ell_j^\mathrm{max}\) sampled at \(2 \ell_j^\mathrm{max}\)), requiring additional spherical harmonic transforms to double the resolution. Also, our spherical harmonic transform function does not calculate harmonic coefficients to a multipole less than the band-limit of the input map (\ie to only calculate \(a_{\ell m}\) for \(\ell < L\) where \(L < \ell_\mathrm{max}^j\)), resulting in excess computation at certain steps in the algorithm. These optimisations are left as further work.

\section{Sources of error in the ILC}
\label{sec:ilc_error}

By the linearity of the wavelet transform in Eq.~\eqref{eq:wav_coeffs}, the data model in Eqs.~\eqref{eq:model} and \eqref{eq:signal} can be recast in wavelet space as
\begin{equation}\label{eq:wav_model}
W_{jnk}^c = a^c W_{jnk}^{\mathrm{CMB}} + W_{jnk}^{\mathrm{FG},c} + W_{jnk}^{\mathrm{N},c} \, ,
\end{equation}
where \(W_{jnk}^{\mathrm{CMB}}\), \(W_{jnk}^{\mathrm{FG},c}\) and \(W_{jnk}^{\mathrm{N},c}\) are respectively the CMB, foreground and instrumental noise contributions to each wavelet coefficient map. The beams within each component have been absorbed into the component wavelet coefficient maps. Substituting Eq.~\eqref{eq:wav_model} into Eq.~\eqref{eq:ilc_sum} gives
\begin{equation}\label{eq:ilc_expand}
\begin{split}
W_{jnk}^{\mathrm{ILC}} &= \sum_{c=1}^{N_c} a^c \omega_{jnk}^c W_{jnk}^{\mathrm{CMB}} + \sum_{c=1}^{N_c} \omega_{jnk}^c (W_{jnk}^{\mathrm{FG},c} + W_{jnk}^{\mathrm{N},c}) \\
&= W_{jnk}^{\mathrm{CMB}} + \frac{\sum_{c,c'=1}^{N_c} (W_{jnk}^{\mathrm{FG},c} + W_{jnk}^{\mathrm{N},c}) (R_{jnk}^{-1})^{cc'} a^{c'}}{\sum_{c,c'=1}^{N_c} a^c (R_{jnk}^{-1})^{cc'} a^{c'}} \, ,
\end{split}
\end{equation}
where the second equality follows by applying the constraint given in Eq.~\eqref{eq:constraint} and expanding the weights as given in Eq.~\eqref{eq:weights}. Even when the calibration \(a^c\) and the covariance matrices \((R_{jnk})^{cc'}\) are correct, there is always residual signal in the final ILC wavelet coefficient maps, given by the second term on the RHS of Eq.~\eqref{eq:ilc_expand}. Due to the linearity of the inverse wavelet transforms, this residual signal will propagate linearly into the final ILC temperature map as calculated by Eq.~\eqref{eq:wav_syn}. As explained in \S~\ref{sec:ilc}, this error term is reduced by minimising the empirical variance of the ILC map assuming that the CMB and foregrounds and the CMB and noise are respectively uncorrelated.

\begin{figure}
\includegraphics[width=\columnwidth]{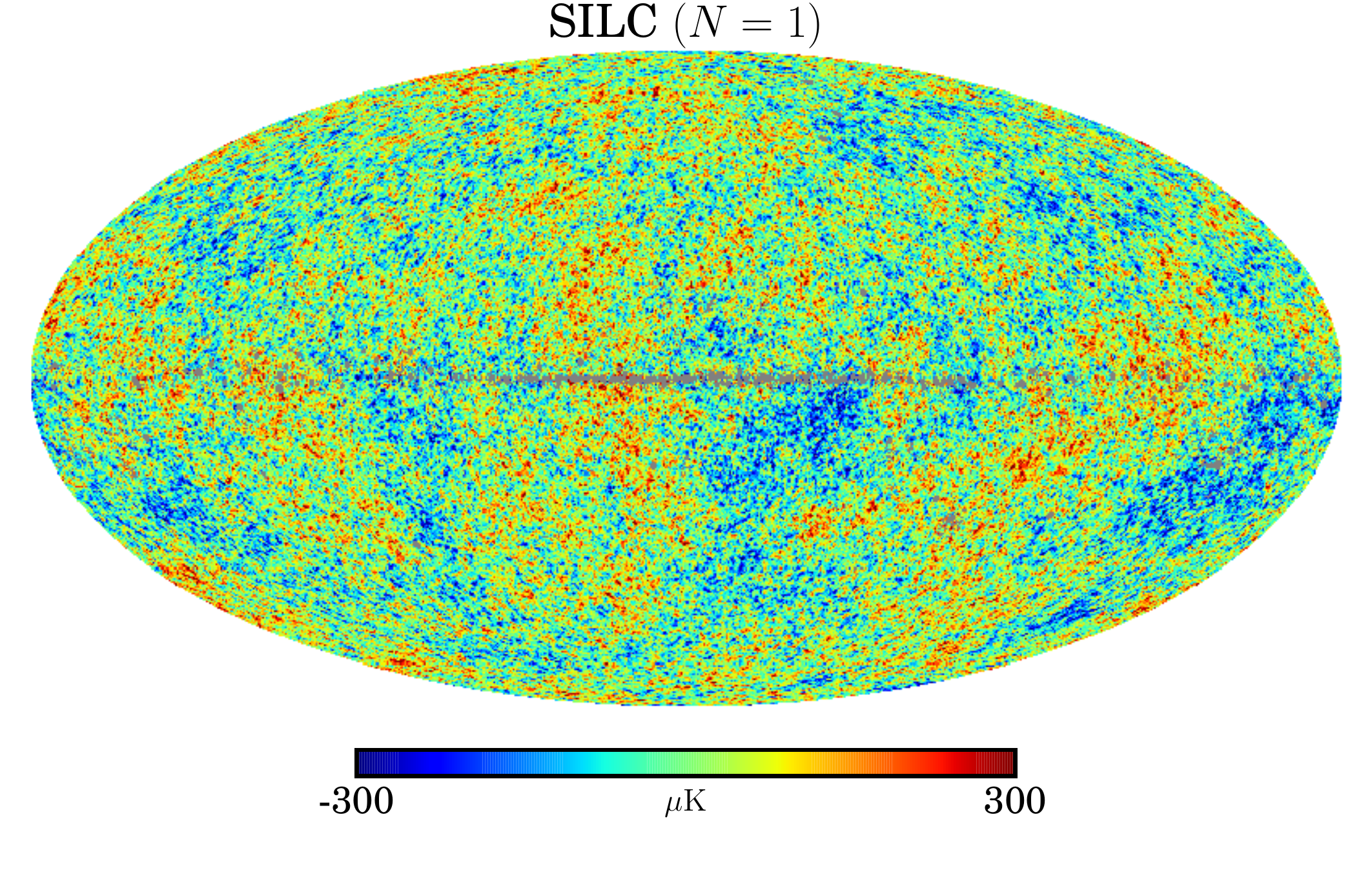} 
\caption{Planck \textit{data.} The CMB temperature anisotropies reconstructed using SILC in the axisymmetric limit \((N = 1, \mathrm{FWHM} = 5\arcmin, N_\mathrm{side} = 2048)\). The grey pixels are the point source mask.}
\label{fig:n1map}
\end{figure}

\begin{figure}
\begin{tabular}{c}
\includegraphics[width=\columnwidth]{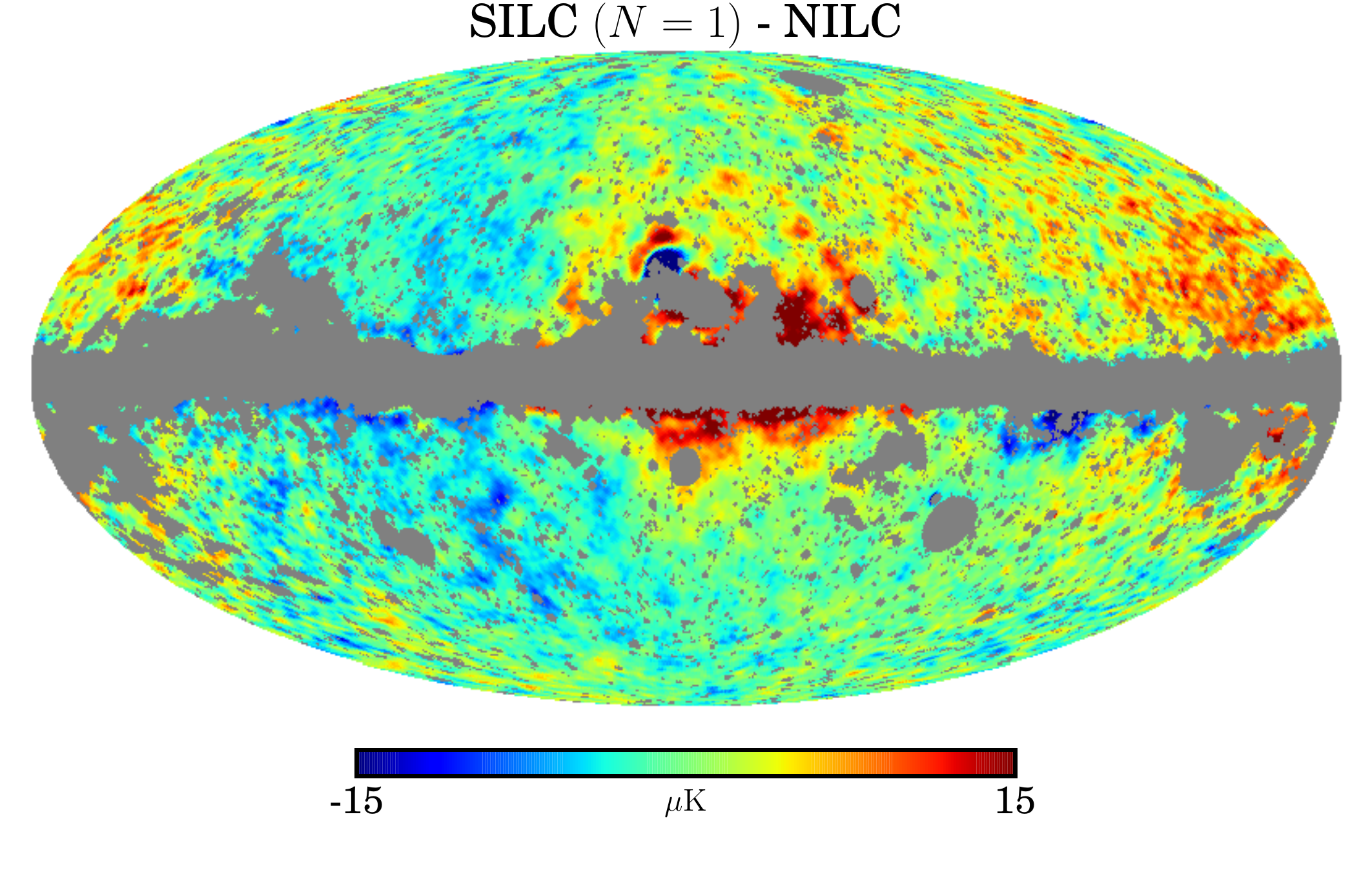} \\
\includegraphics[width=\columnwidth]{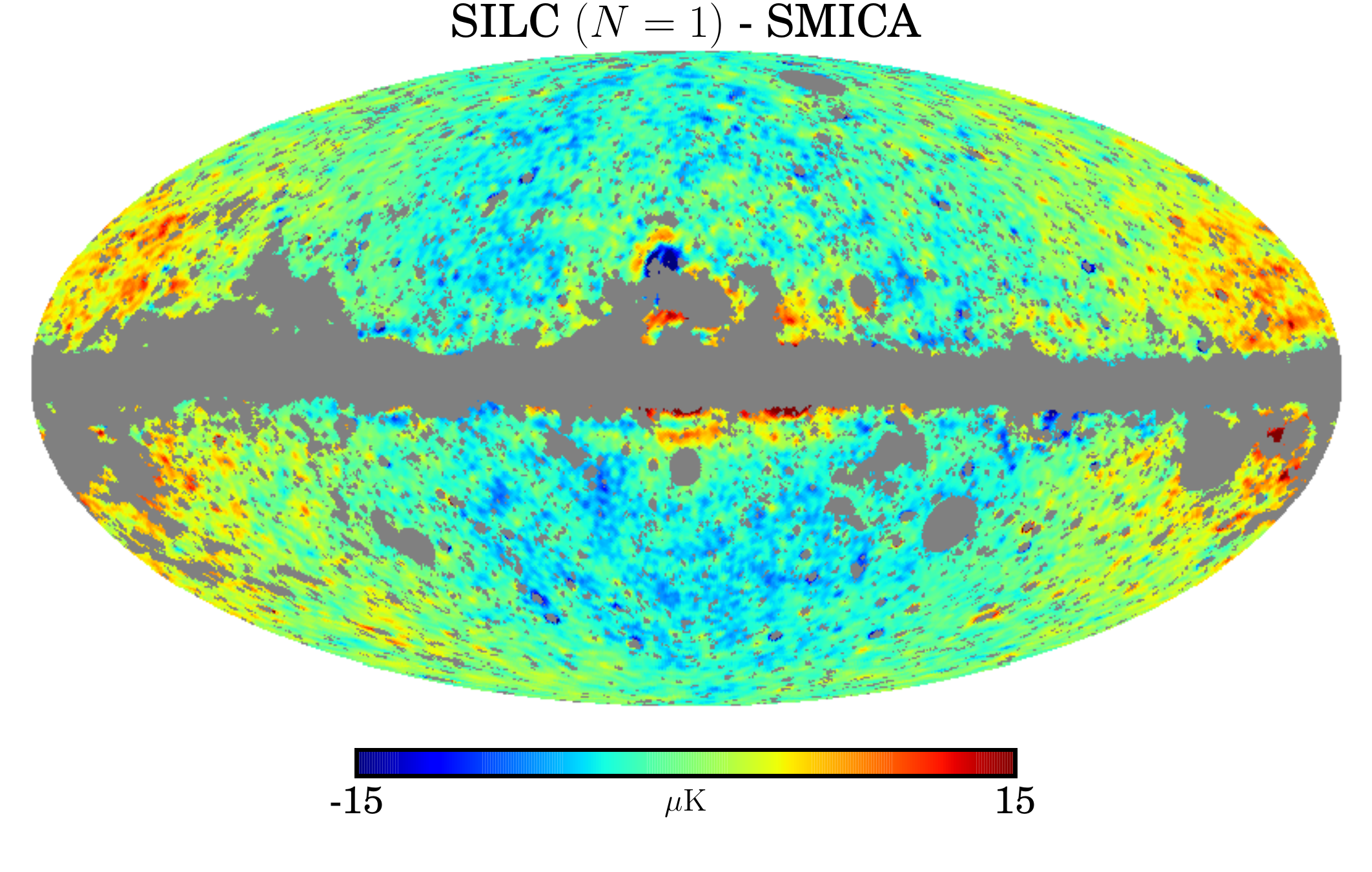} \\
\includegraphics[width=\columnwidth]{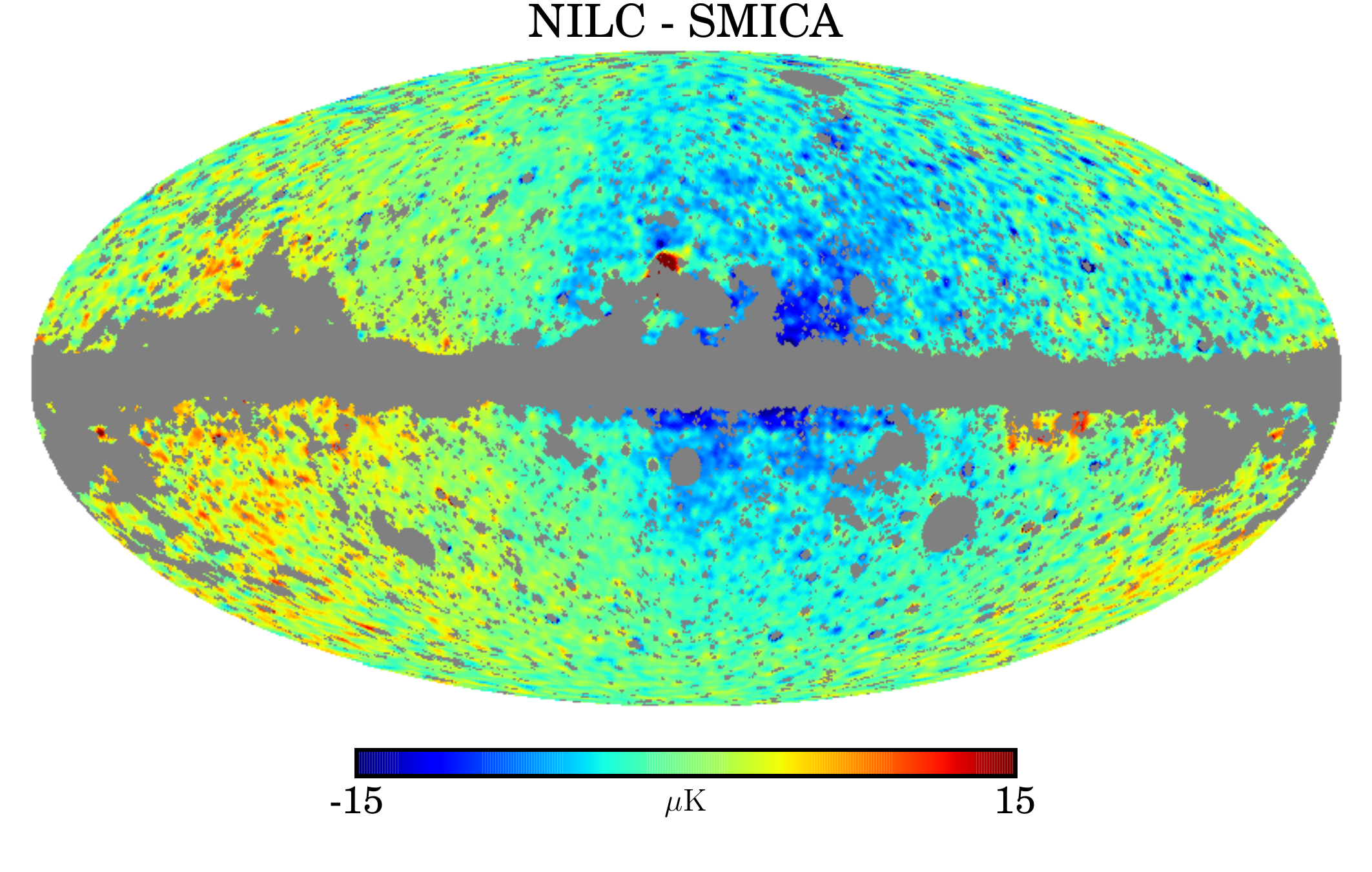}
\end{tabular}
\caption{Planck \textit{data.} Differences between the axisymmetric limit (\(N=1\)) of SILC, NILC and SMICA. The maps have been smoothed to \(\mathrm{FWHM} = 80\arcmin\) and downgraded to \(N_\mathrm{side} = 128\). The grey pixels are the UT78 confidence mask from \citet{2015arXiv150205956P}, which masks the regions of the NILC and SMICA maps not recommended for cosmological analysis. The differences are (\textit{from top to bottom}) (a) SILC (\(N=1\)) - NILC, (b) SILC (\(N=1\)) - SMICA and (c) NILC - SMICA.}
\label{fig:nilcdiffmap}
\end{figure}

There are additional sources of error in the ILC method. The first is due to inaccuracy in the calculation of covariance matrices \((R_{jnk})^{cc'}\), \ie deviations in the empirical estimate \((\hat{R}_{jnk})^{cc'}\) from the true covariance \((R_{jnk})^{cc'}\). \citet{2009A&A...493..835D} estimated the first order expansion of the reconstruction error in the ILC map estimate due to this covariance error. They showed that the covariance of the ILC error with the CMB is inversely proportional to the number of ``effective modes'' used in the ILC calculation. This covariance bias is negative. In our directional wavelet decomposition, our ``effective modes'' are spherical harmonic coefficients weighted to take account of the fact that the harmonic responses of wavelets overlap in both scale and direction. As \(N\), the number of orientations probed, increases and so does the number of wavelets, each wavelet coefficient map contains fewer ``effective modes.'' We therefore conclude that our directional wavelet ILC may be susceptible to this negative ILC bias by increasing \(N\). \citet{2009A&A...493..835D} also showed that due to chance correlations between the CMB and foregrounds, the variance minimisation leads to the unintentional cancellation of \(N^c - 1\) CMB modes. For {\Planck}, \(N^c = 9\), whereas for {\WMAP}, \(N^c = 5\). We therefore expect the magnitude of this negative bias to double simply by using more input frequency channels. Also, since this covariance bias is due to the cancellation of CMB modes, \citet{2009A&A...493..835D} showed that the absolute value is proportional to the CMB power. Therefore, the absolute value of the bias is greatest on large scales where CMB power is concentrated. In general, these biases are best estimated through suites of Monte Carlo simulations.

Another source of error is due to inaccuracy in the calibration \(a^c\) of the CMB. \citet{2010MNRAS.401.1602D} calculated the consequence of a first order error in \(a^c\) on a multiplicative correction to the CMB term in Eq.~\eqref{eq:ilc_expand}. They showed that even a small error in calibration can lead to a significant negative multiplicative bias in the CMB term, when the signal-to-noise ratio is large. (Here, the noise in this ratio also includes foreground signal.) They consider the implications for using an ILC on {\Planck} data, where the signal-to-noise ratio is larger than for {\WMAP} data. They estimate that a 1\% error in \(a^c\) can cancel about a third of the CMB signal, while even a 0.1\% error in \(a^c\) can remove about 1\% of the CMB. Since our main map products use {\Planck} data as input, they will be susceptible to this additional negative calibration bias. In this work, we assume that the CMB is calibrated to have unit response for all frequency channels, \ie \(a^c = 1\) for all \(c\).

As mentioned in \S~\ref{sec:model} and \ref{sec:beam}, we assume all beams to be circularly symmetric. Therefore, non-axisymmetric beam components will propagate into the ILC calculation but are assumed sufficiently small to be ignored.

\section{Comparison to previous work}
\label{sec:comp}

\renewcommand{\arraystretch}{0}
\begin{figure}
\begin{tabular}{c}
\includegraphics[width=\columnwidth]{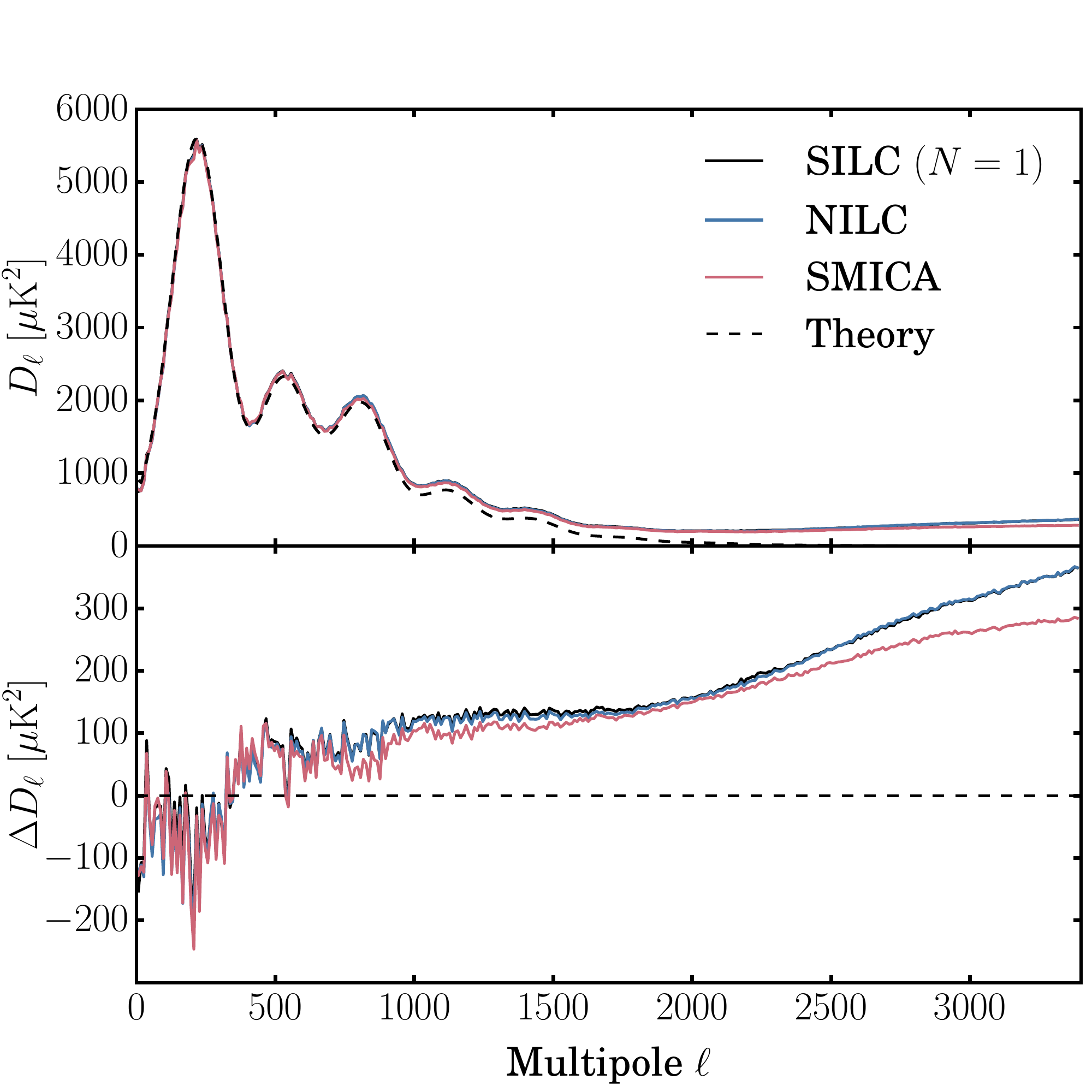} \\
\includegraphics[width=\columnwidth]{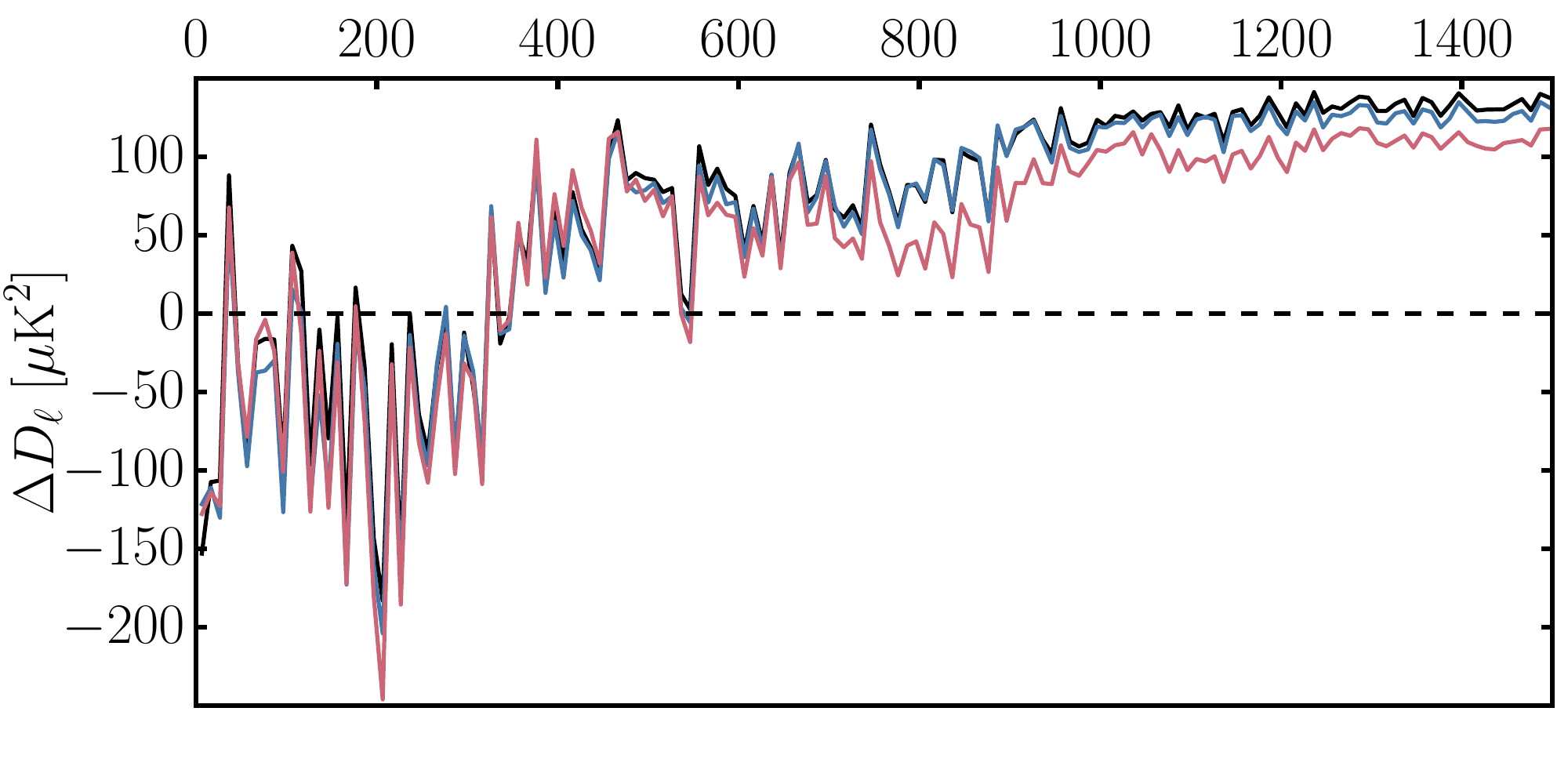}
\end{tabular}
\caption{Planck \textit{data.} \(TT\) angular power spectra comparing the axisymmetric limit \((N = 1)\) of SILC to NILC and SMICA. The top panel (a) shows point source masked spectra. The middle panel (b) shows residuals after subtracting the best-fit \(\Lambda\)CDM model from the {\Planck} 2015 likelihood. The bottom panel (c) shows the same residuals at low multipoles only (\(\ell < 1500\)).}
\label{fig:nilcspec}
\end{figure}
\renewcommand{\arraystretch}{1}

We now consider how SILC compares with existing component separation methods, particularly those adopted for the {\Planck} analysis. We applied the axisymmetric limit (when \(N = 1\)) of SILC to full-mission {\Planck} data and compared the results to existing {\Planck} analyses using the NILC and SMICA methods: the former because it is the closest in spirit to SILC, and the latter because it is the baseline method adopted by the Planck Collaboration for high-resolution analyses. Figure \ref{fig:n1map} shows the CMB reconstructed by the axisymmetric limit of SILC, while Fig.~\ref{fig:nilcdiffmap} shows the differences between this map and the NILC and SMICA (full-mission 2015 release) CMB maps and the difference between NILC and SMICA. The differences between the three maps are small in magnitude and mostly concentrated at the edges of the Galactic mask towards the Galactic centre, where foreground emission is most intense and complex. Quantitatively, we can compare the mean values and standard deviations of the masked difference maps. The mean values of Figs.~\ref{fig:nilcdiffmap} (a), (b) and (c) are respectively $0.44$, $-0.63$, and $-1.07$ \(\mu \mathrm{K}\), while the standard deviations are respectively $4.24$, $3.38$ and $3.43$ \({\mu \mathrm{K}}^2\). These values are small and similar, suggesting a strong consistency between the three methods. These difference maps have been formed from maps which have been smoothed and downgraded in resolution and so visually highlight differences at the lowest multipoles.

\renewcommand{\arraystretch}{0}
\begin{figure}
\begin{tabular}{c}
\includegraphics[width=\columnwidth]{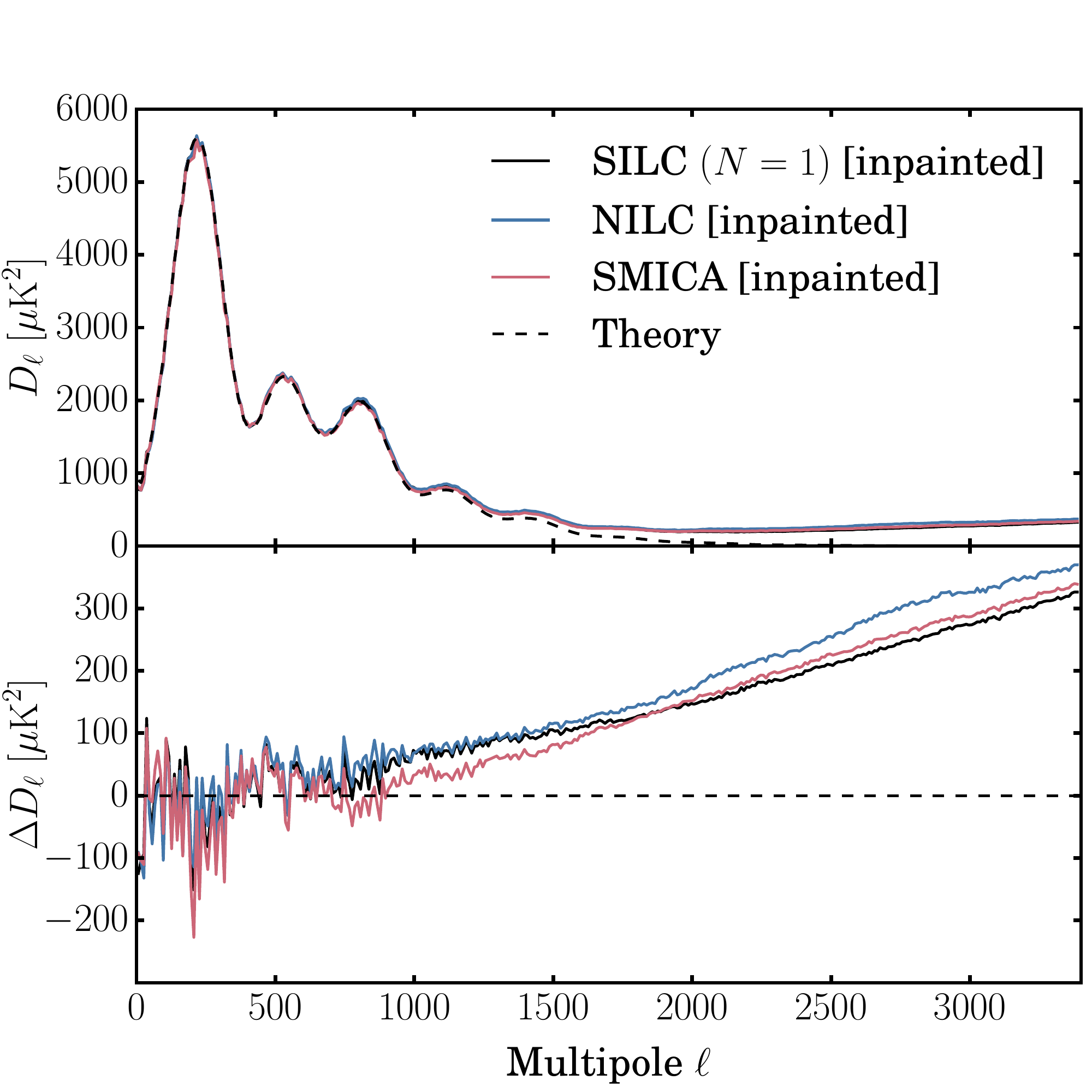} \\
\includegraphics[width=\columnwidth]{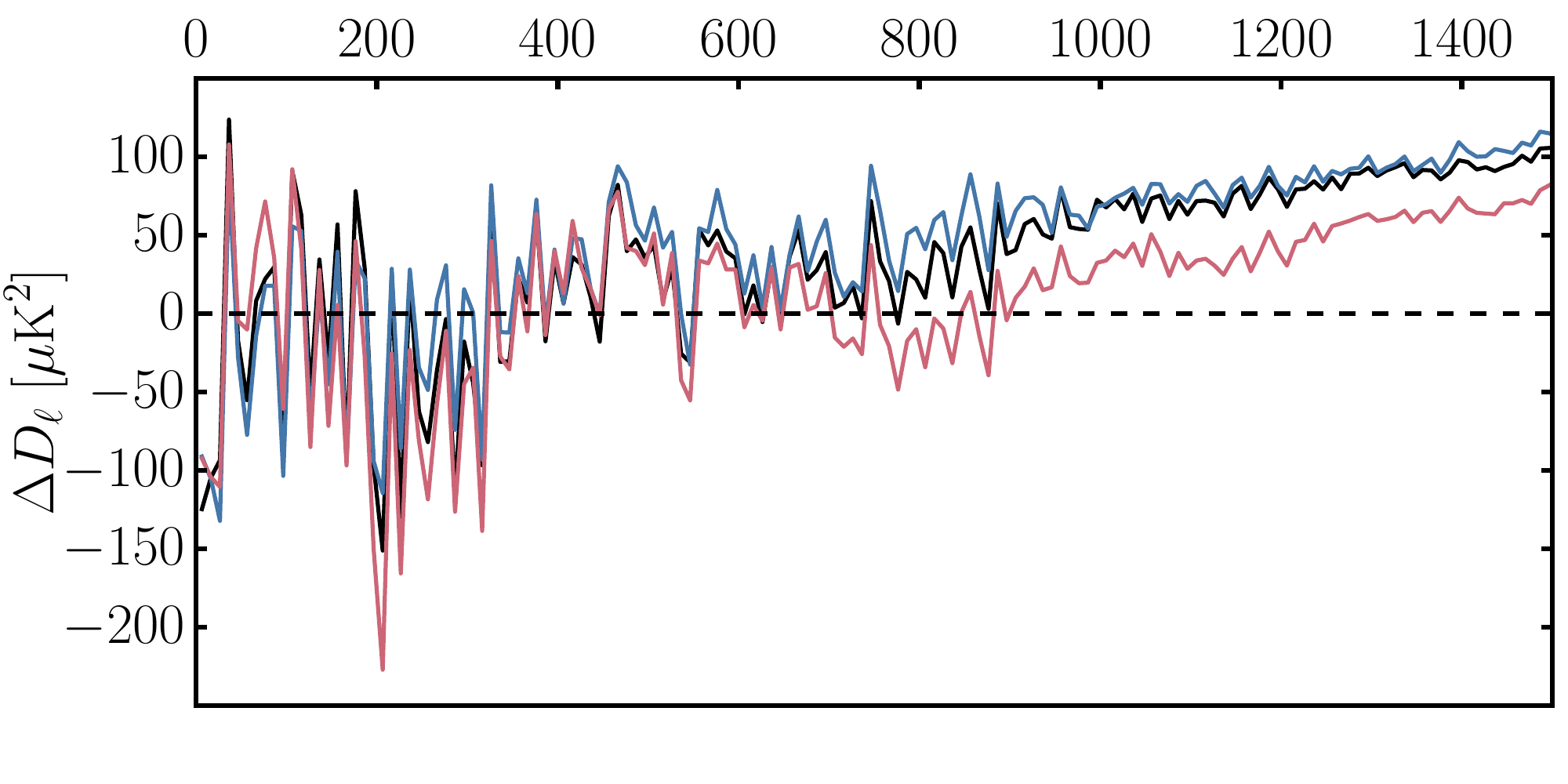}
\end{tabular}
\caption{Planck \textit{data.} \(TT\) angular power spectra comparing the axisymmetric limit \((N = 1)\) of SILC to NILC and SMICA. The top panel (a) shows full-sky spectra of inpainted maps. The middle panel (b) shows residuals after subtracting the best-fit \(\Lambda\)CDM model from the {\Planck} 2015 likelihood. The bottom panel (c) shows the same residuals at low multipoles only (\(\ell < 1500\)).}
\label{fig:paintspec}
\end{figure}
\renewcommand{\arraystretch}{1}

Figure \ref{fig:nilcspec} compares point source masked \(TT\) angular power spectra \((D_\ell = \ell (\ell+1) C_\ell / 2 \pi)\) at the full multipole range of the three maps (up to \(\ell = 3400\))\footnote{In order to estimate full-sky spectra from a masked map, we correct the \(C_\ell\) by dividing by \(f_\mathrm{sky} = 0.978\), a good approximation for a small mask. We elect to use point source masked spectra in order to concentrate our analysis on foreground and noise removal, rather than how maps are inpainted; all three maps are inpainted (at least) within the mask used.} with a CMB spectrum derived from the {\Planck} 2015 \(TT\) and low \(TEB\) likelihood\footnote{The parameters come from the \texttt{base\_plikHM\_TT\_lowTEB} likelihood. The values are available in the {\Planck} 2015 Release Explanatory Supplement: 2015 Cosmological parameters and MC chains (\url{http://wiki.cosmos.esa.int/planckpla2015/images/f/f7/Baseline_params_table_2015_limit68.pdf}).}. The SILC spectrum is remarkably similar to that of NILC. This is unsurprising since the axisymmetric limit of SILC (when \(N = 1\)) is very similar to the NILC method. Nonetheless, there are a number of pipeline differences. In particular, we use a different set of wavelets than the needlets employed in NILC (as discussed in \S~\ref{sec:wavelets}), even in the axisymmetric limit, with different harmonic responses. Figure \ref{fig:wavelets} shows the harmonic response of the wavelets used in this work and Table \ref{tab:wav_lims} lists their harmonic band-limits \(\ell^j_{\mathrm{min}}\) and \(\ell^j_{\mathrm{max}}\). The SMICA spectrum has lower residuals at higher multipoles than both the axisymmetric limit of SILC and NILC.

\begin{figure}
\includegraphics[width=\columnwidth]{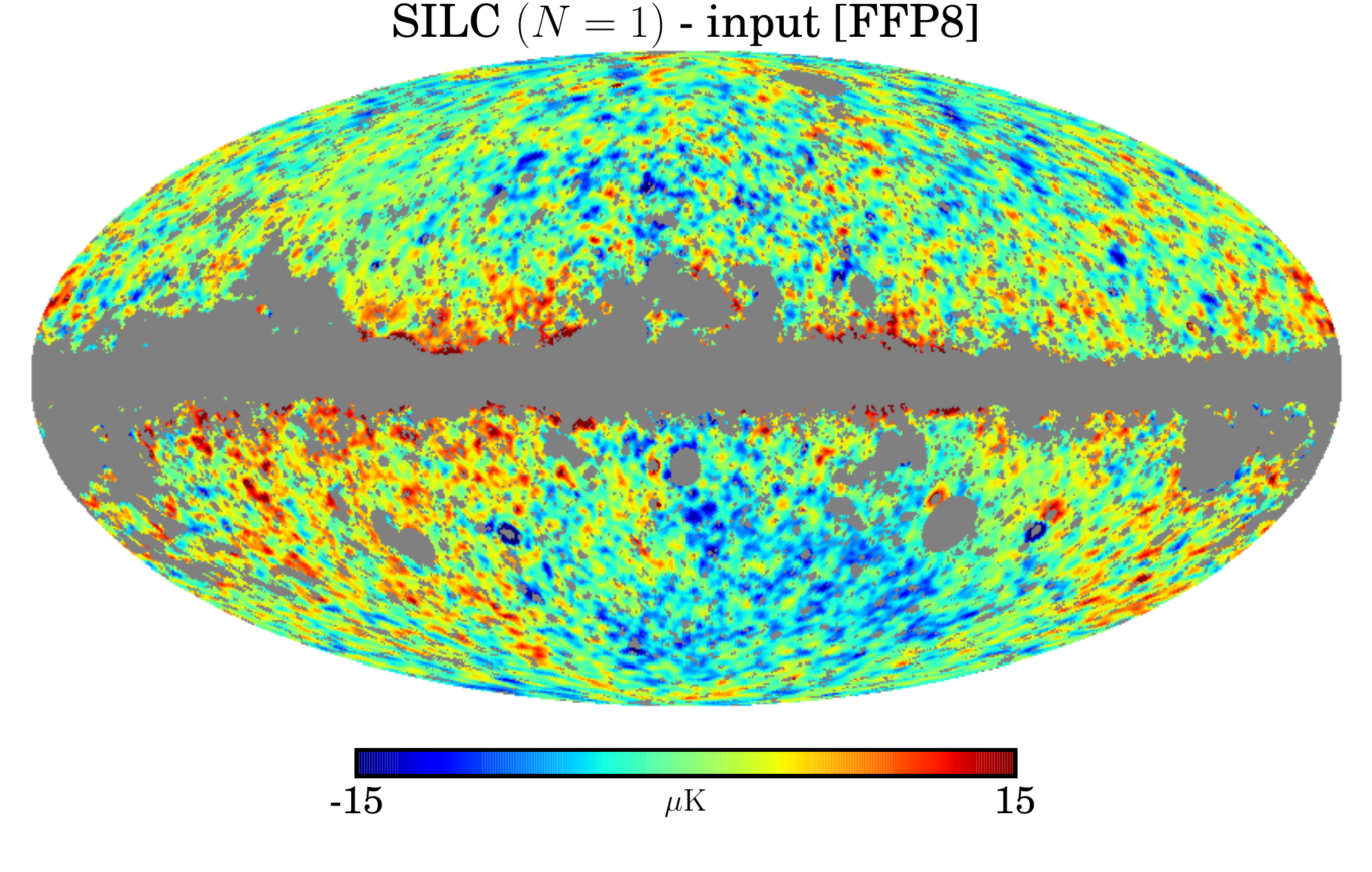}
\caption{Planck \textit{simulations.} Difference between output ILC and input CMB temperature maps from FFP8 simulations. The maps have been smoothed to \(\mathrm{FWHM} = 80\arcmin\) and downgraded to \(N_\mathrm{side} = 128\). The grey pixels are the UTA76 confidence mask from \citet{2015arXiv150205956P}, which masks the Galactic region in FFP8 simulations where foreground emission is strongest.}
\label{fig:ffp8diffmap}
\end{figure}

Figure~\ref{fig:paintspec} compares full-sky angular power spectra of the three maps, including the inpainted point source regions. The spectra are similar to those in Fig.~\ref{fig:nilcspec}. The main difference is the lower noise tail in the SILC map at high multipoles above \(\ell = 1500\) (where all component separation CMB maps are dominated by residual instrumental noise). This is because, unlike NILC and SMICA, we do not Gaussian inpaint the very largest point source holes, but rather use diffusive inpainting (as discussed in \S~\ref{sec:ps}). The Gaussian inpainting of large irregular holes is poorly constrained and adds residual noise relative to diffusive inpainting. 

We have shown that the axisymmetric limit of SILC gives comparable performance to NILC and SMICA. In \S~\ref{sec:simulations} and \S~\ref{sec:data}, we ``turn on'' the directionality of the wavelets and consider the impact on CMB reconstruction from simulated and real data respectively.

\section{Application to \textit{Planck} simulations}
\label{sec:simulations}

\begin{figure} 
\begin{tabular}{c}
\includegraphics[width=\columnwidth]{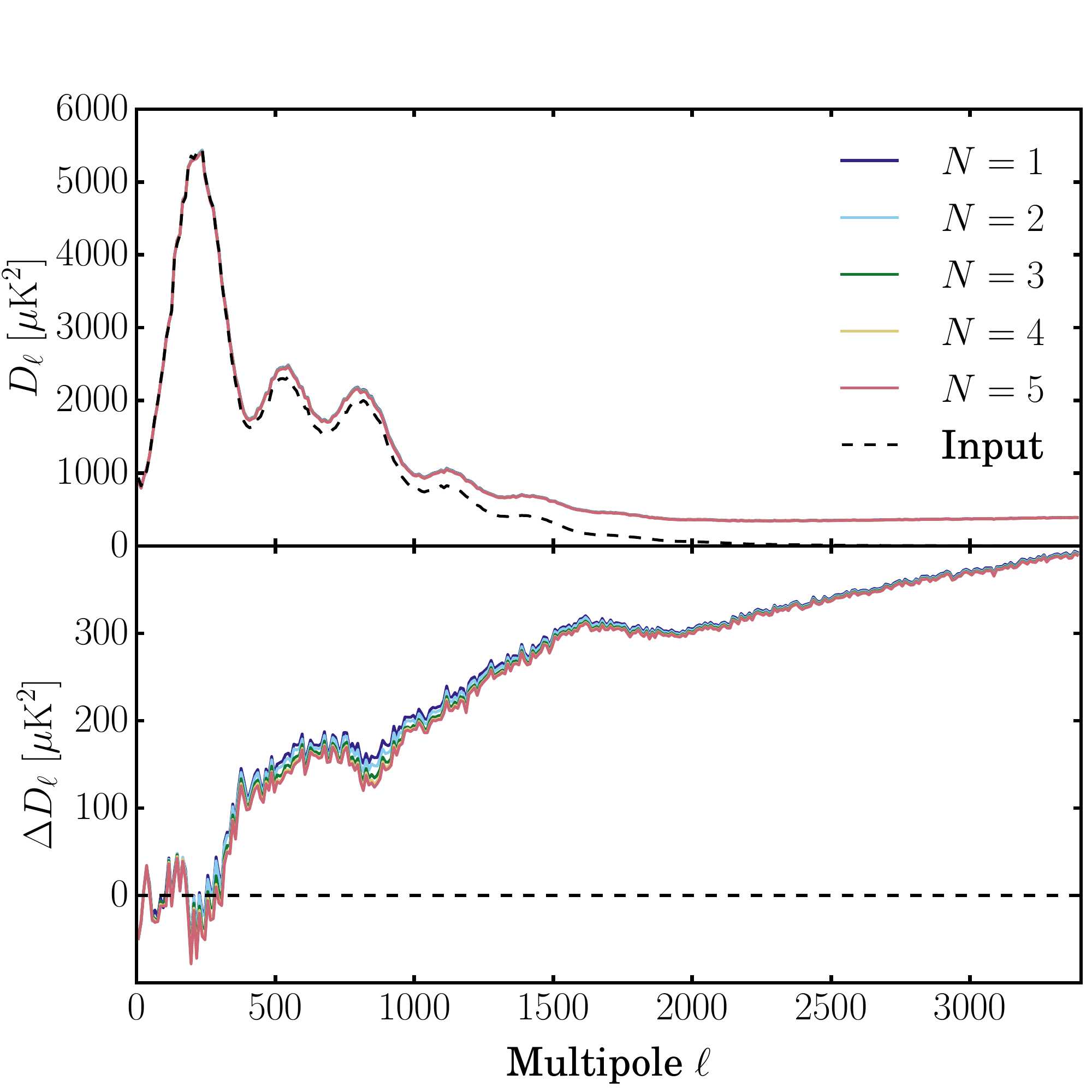} \\
\includegraphics[width=\columnwidth]{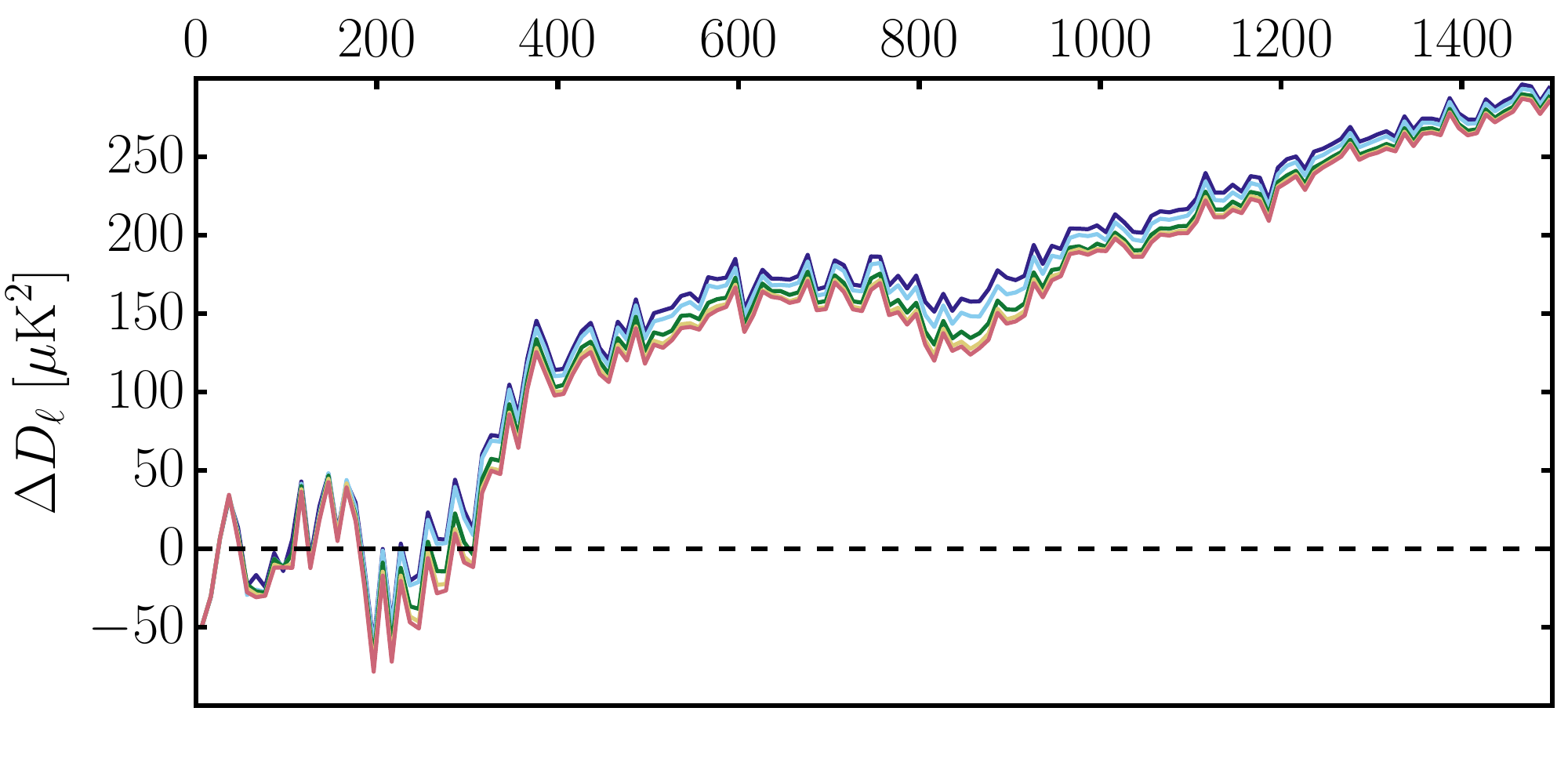}
\end{tabular}
\caption{Planck \textit{simulations.} \(TT\) angular power spectra comparing output ILC using different values of \(N\) and input CMB from FFP8 simulations. The top panel (a) shows point source masked spectra. The middle panel (b) shows residuals after subtracting the input CMB spectrum. The bottom panel (c) shows the same residuals at low multipoles only (\(\ell < 1500\)).}
\label{fig:ffp8spec}
\end{figure}

\begin{figure*}
\begin{tabular}{cc}
\includegraphics[width=\columnwidth]{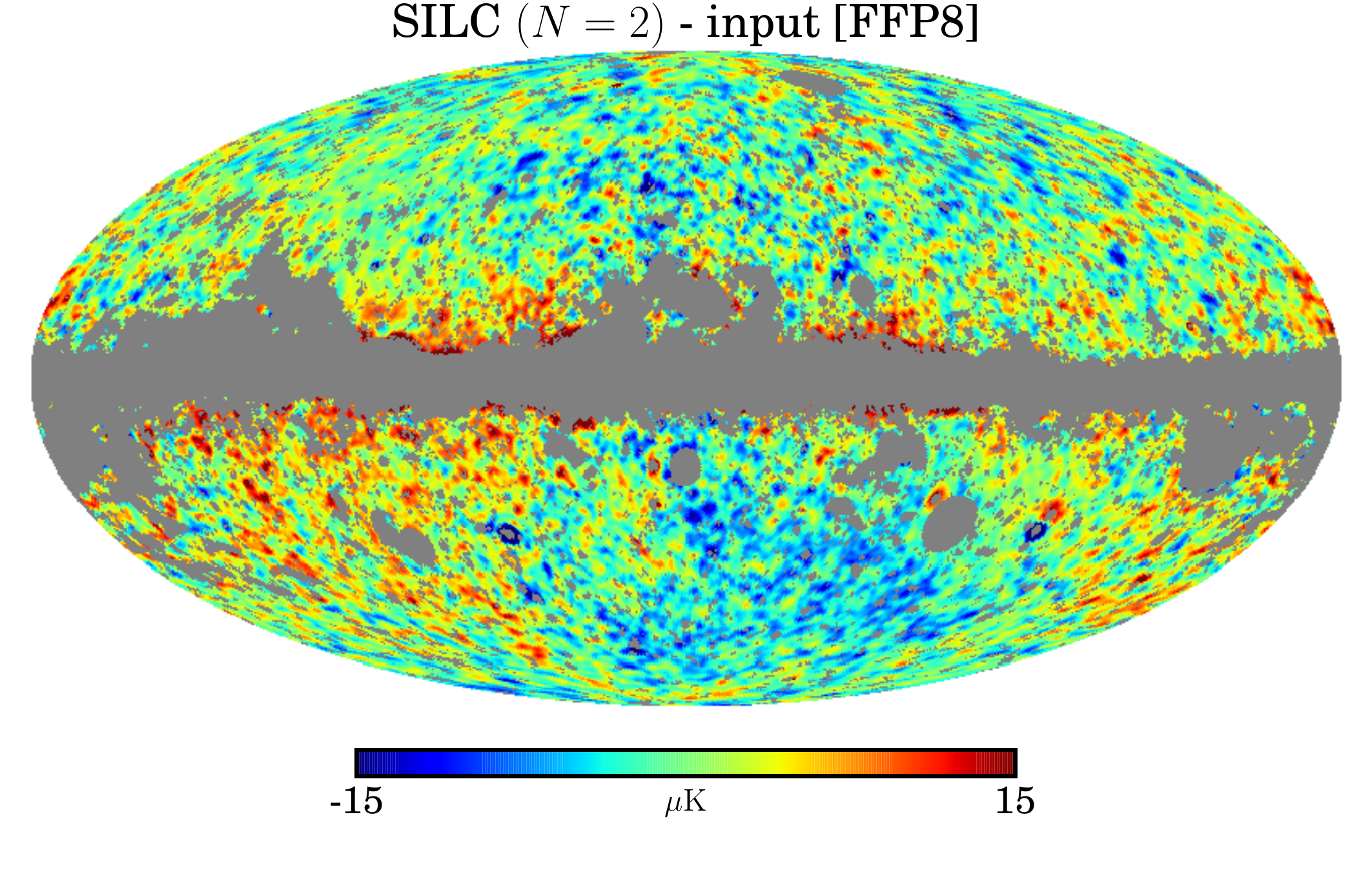} & \includegraphics[width=\columnwidth]{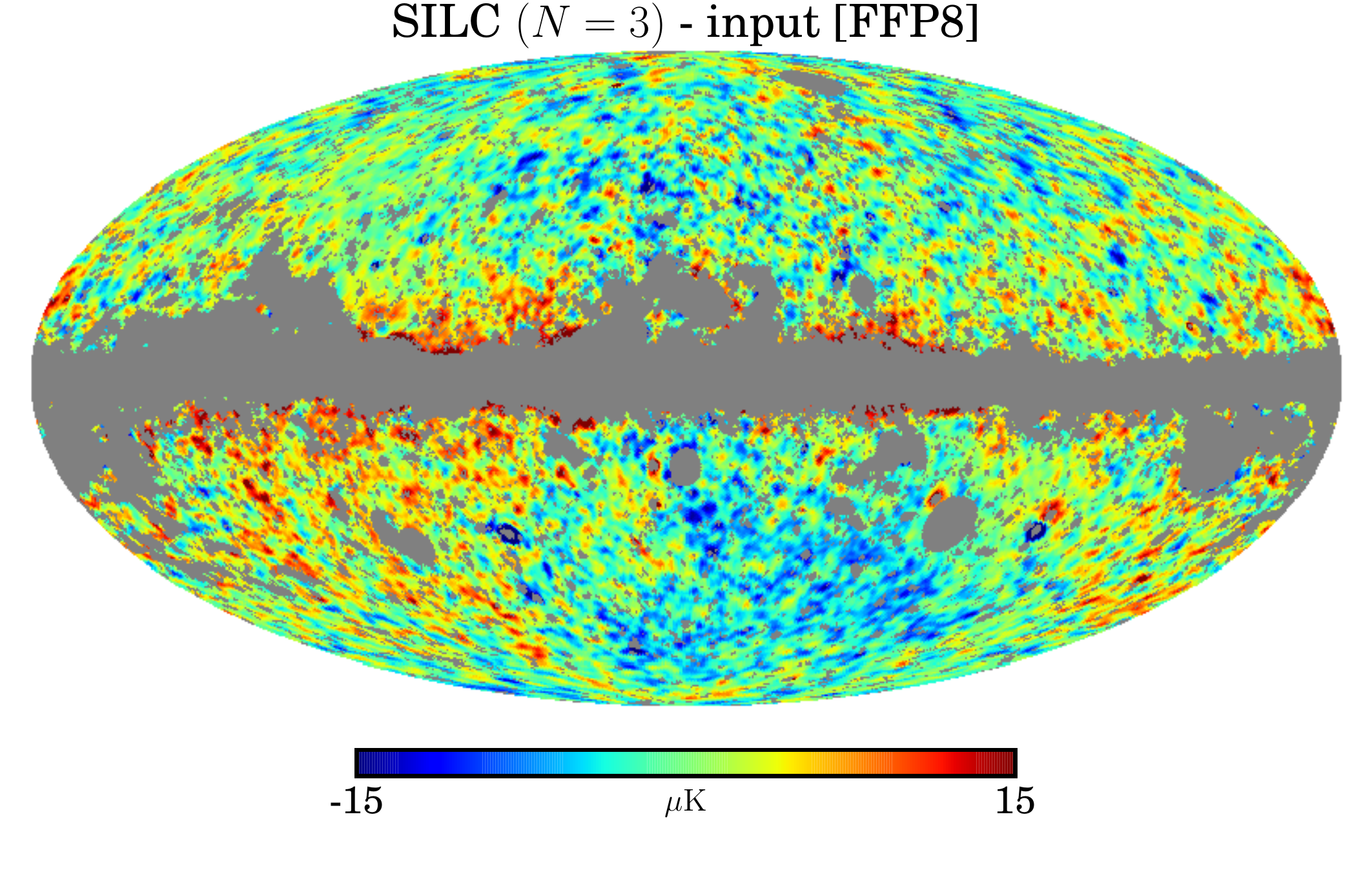} \\
\includegraphics[width=\columnwidth]{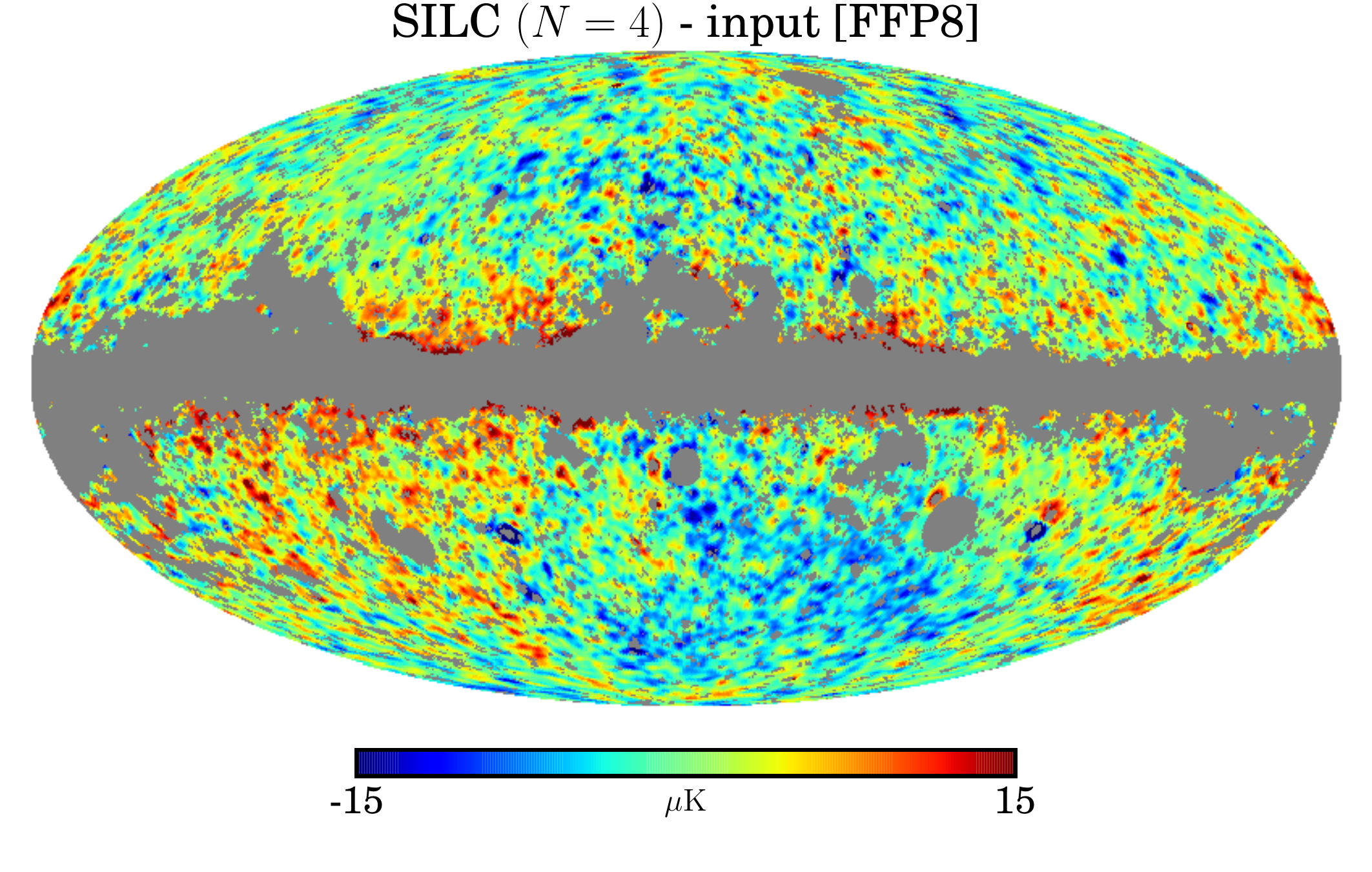} & \includegraphics[width=\columnwidth]{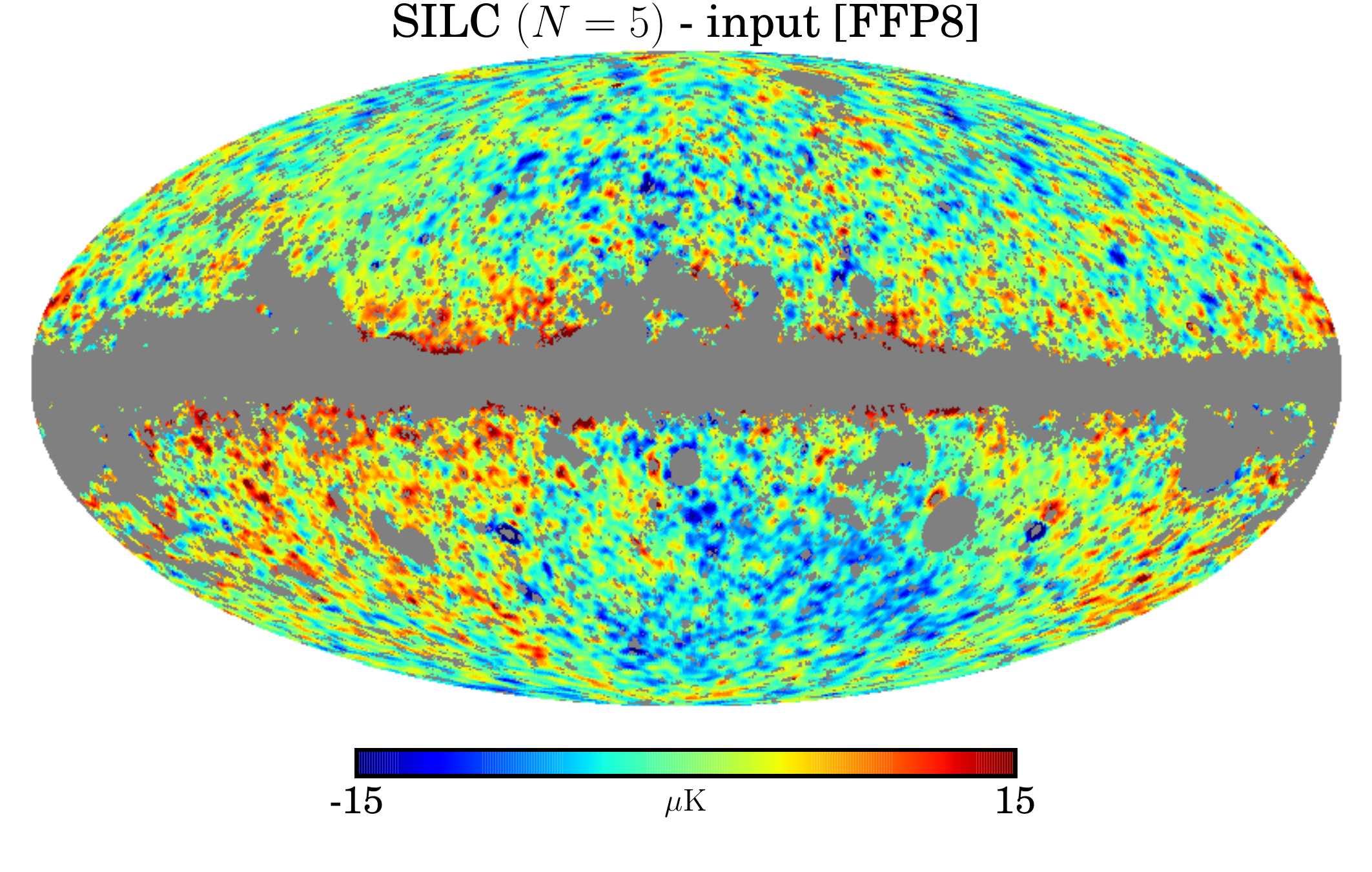}
\end{tabular}
\caption{Planck \textit{simulations.} Differences between output ILC reconstructed using different values of \(N\) and input CMB temperature maps from FFP8 simulations. The maps have been smoothed to \(\mathrm{FWHM} = 80\arcmin\) and downgraded to \(N_\mathrm{side} = 128\). The grey pixels are the UTA76 confidence mask. The differences are (\textit{from left to right, top to bottom}) (a) \(N = 2\), (b) \(N = 3\), (c) \(N = 4\), (d) \(N = 5\) minus the input CMB.}
\label{fig:ffp8_n5diffmap}
\end{figure*}

We now apply SILC to the fiducial full-mission {\Planck} FFP8 simulated sky maps, focusing on the impact on CMB reconstruction by increasing directionality as a function of scale. Figure \ref{fig:ffp8diffmap} shows the difference between our reconstructed CMB (using \(N = 1\)) and the input simulated CMB. There are small-magnitude differences particularly at the edge of the Galactic mask where the strength and complexity of foreground emission is greatest. As in Fig.~\ref{fig:nilcdiffmap}, this difference map is at low resolution and so highlights residuals at the lowest multipoles. Figure \ref{fig:ffp8spec} compares point source masked \(TT\) angular power spectra (up to \(\ell = 3400\)) of CMB maps reconstructed using values of \(N\) from 1 to 5. It can be seen that the introduction of directionality has the greatest effect at multipoles around \(\ell = 800\); the residuals are beginning to converge for \(\ell \goa 2000\). Figure \ref{fig:ffp8_n5diffmap} shows the differences between simulated CMB maps reconstructed using \(N = [2,3,4,5]\) minus the input CMB. The four maps and the axisymmetric difference map in Fig.~\ref{fig:ffp8diffmap} are almost identical with small magnitude residuals. This is because these low-resolution difference maps again highlight residuals on the very largest scales. However, as discussed in \S~\ref{sec:wavelets}, the wavelets we use are constructed to have an axisymmetric scaling function at the very lowest multipoles. The scaling function we use (as detailed in Table \ref{tab:wav_lims}) means that no directionality is applied for \(\ell < 32\).

\renewcommand{\arraystretch}{0}
\renewcommand{\arraystretch}{1}

\begin{figure*}
\begin{tabular}{cc}
\includegraphics[width=\columnwidth]{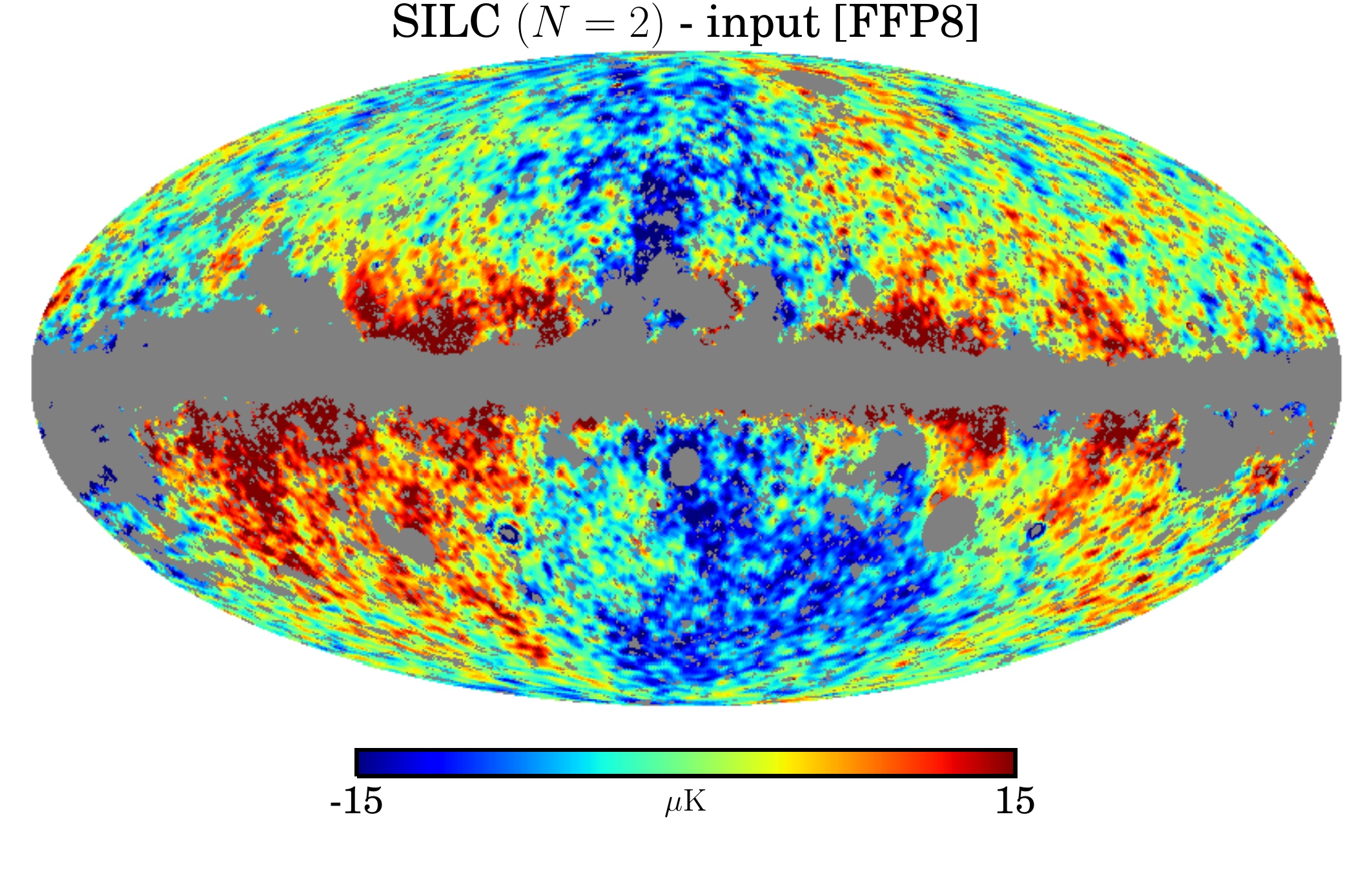} & \includegraphics[width=\columnwidth]{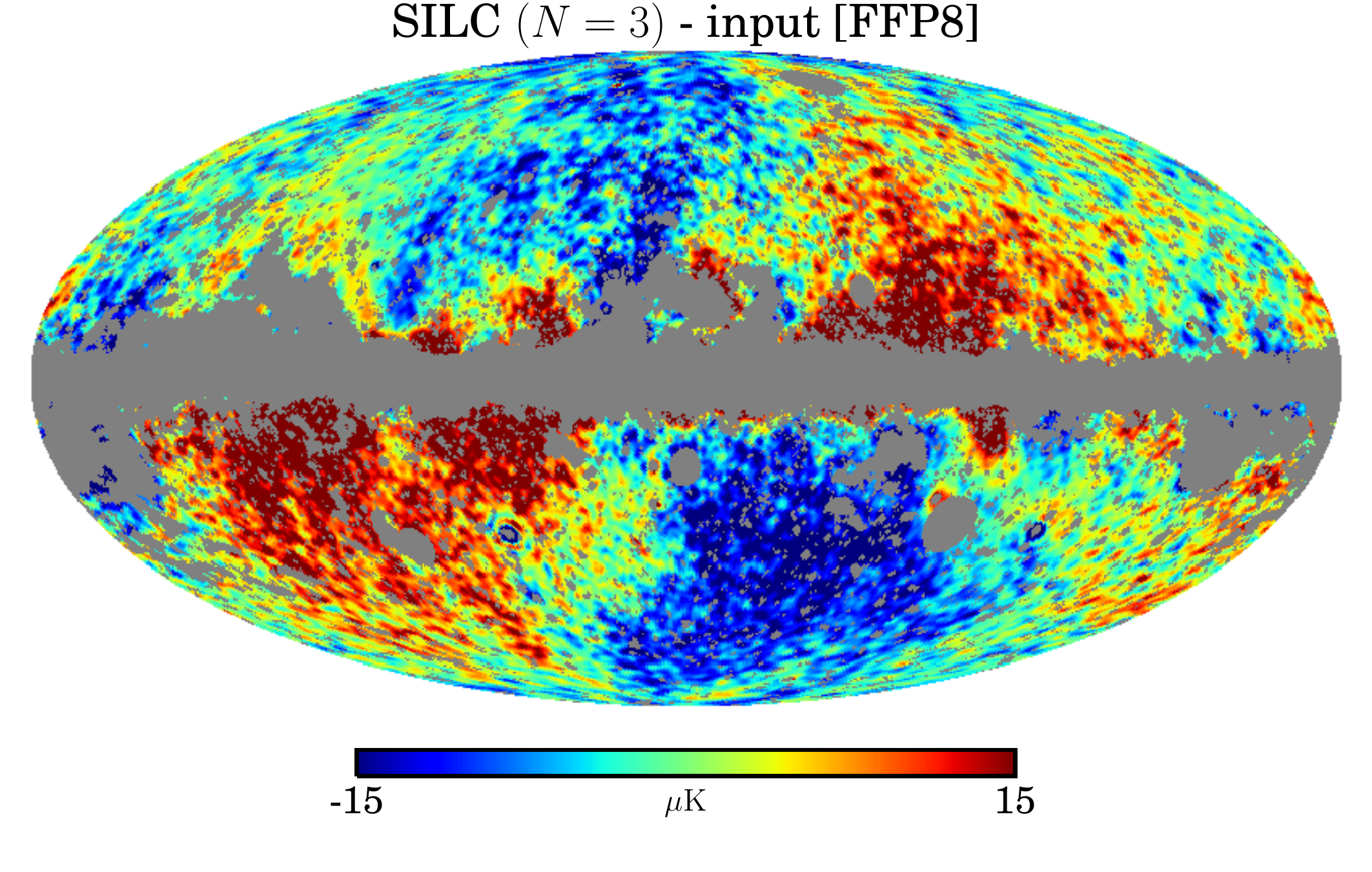} \\
\includegraphics[width=\columnwidth]{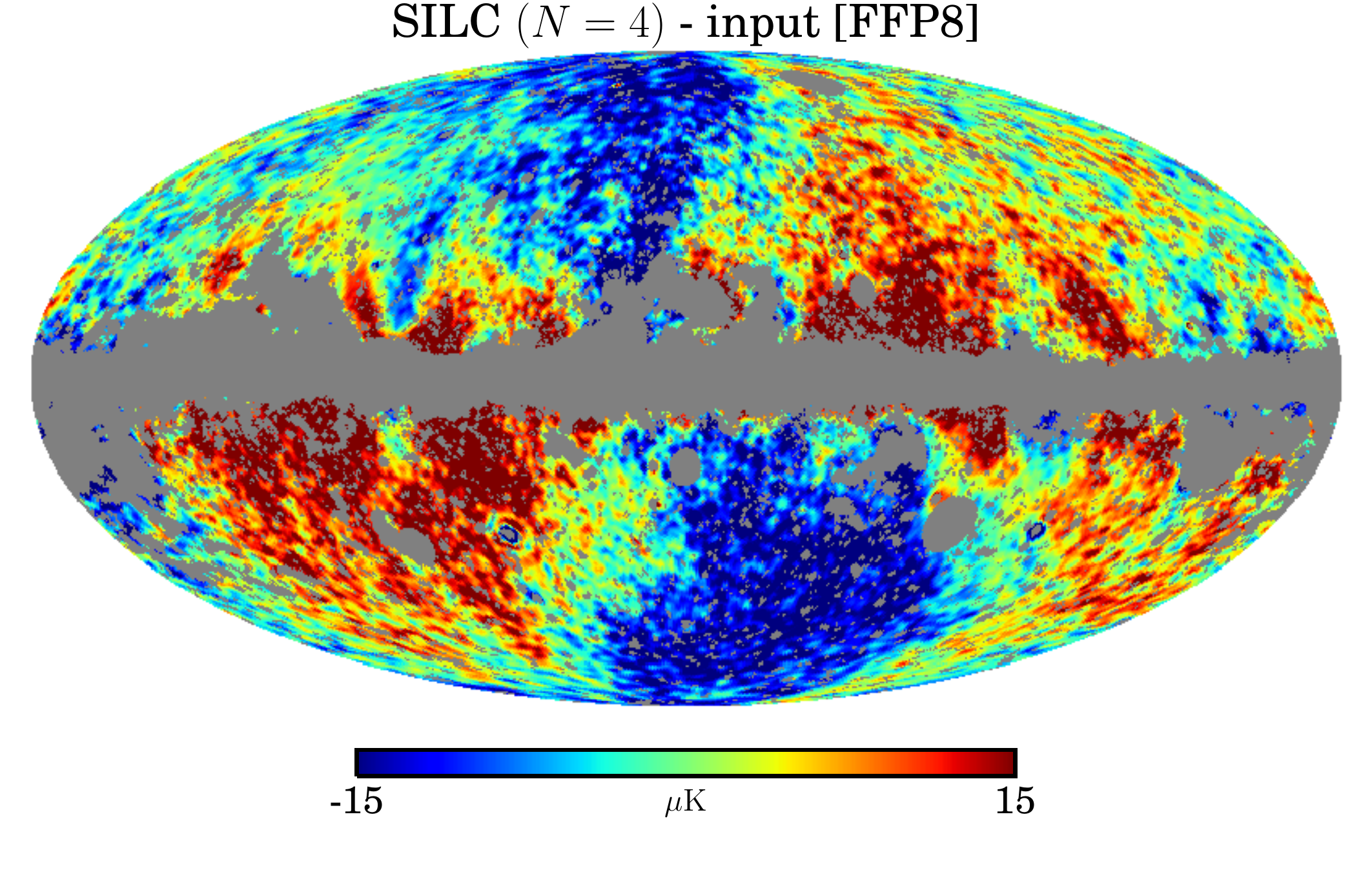} & \includegraphics[width=\columnwidth]{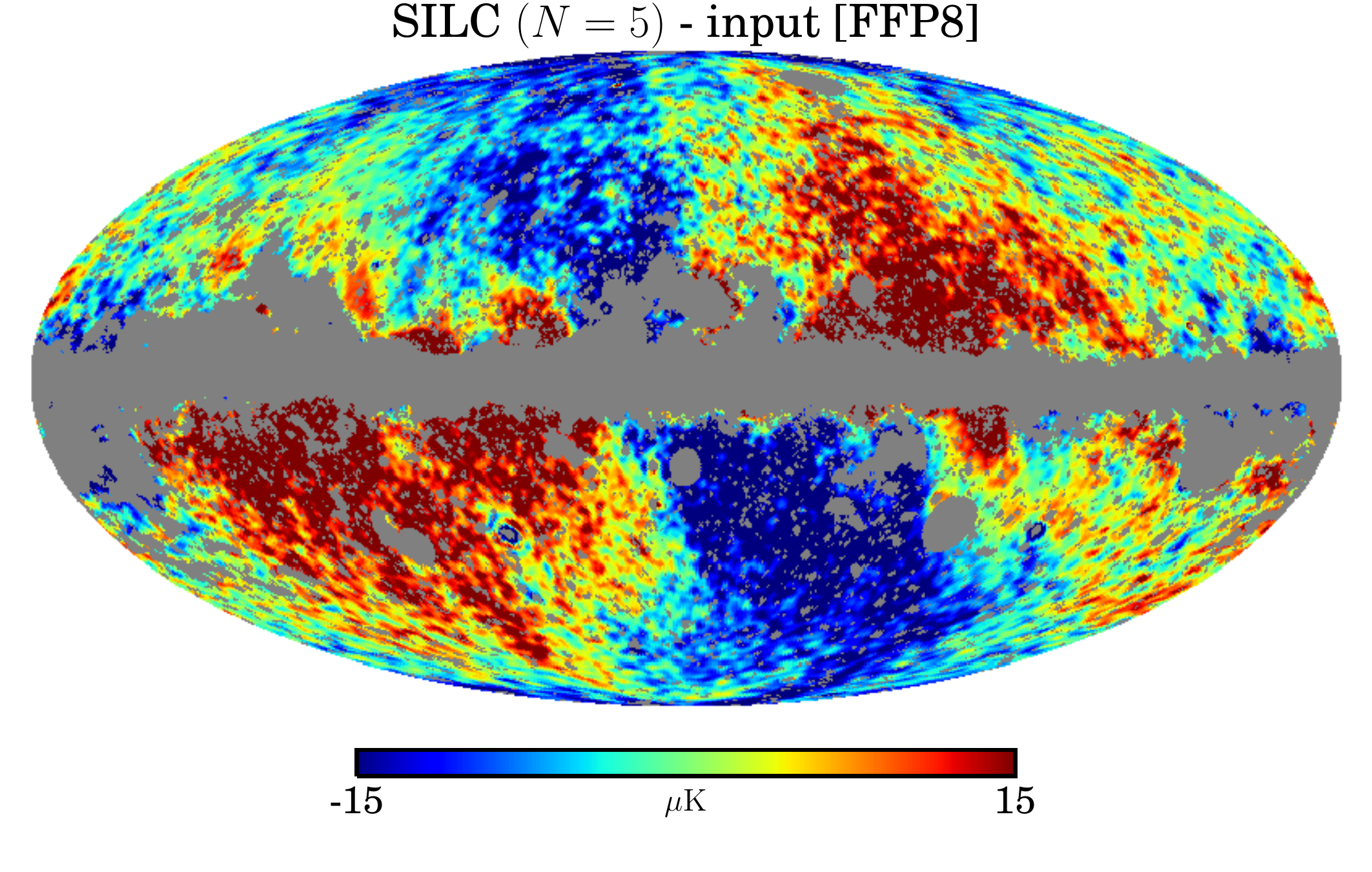}
\end{tabular}
\caption{Planck \textit{simulations.} Same as Fig.~\ref{fig:ffp8_n5diffmap} (which uses the recommended wavelets) but here using directional wavelets on large scales (\(\ell < 32\)), which is not recommended as it leads to increased CMB reconstruction errors as seen above.}
\label{fig:ffp8_n5diffmap_biased}
\end{figure*}

Figure \ref{fig:ffp8_n5diffmap_biased} shows equivalent difference maps as in Fig.~\ref{fig:ffp8_n5diffmap} but for the simulated CMB reconstructed using directional wavelets at all scales\footnote{In particular, the scaling function and \(j = 0\) wavelet are replaced by two directional wavelets with harmonic band-limits \([1,60]\) and \([1,128]\).}, including for \(\ell < 32\). It can be seen that the reconstruction errors are significantly larger in magnitude and cover almost the entire sky. The errors are also dominated by the largest scales, in particular a large error in the quadrupole increasing with magnitude as the amount of directionality \(N\) increases. We attribute this effect most probably to the ILC ``biases'' discussed in \S~\ref{sec:ilc_error}, in particular the cancellation of CMB modes due to chance correlations with foregrounds in the ILC variance minimisation. \citet{2009A&A...493..835D} showed that the absolute value of this effect is largest on large scales where CMB power is concentrated, since the cancelled CMB modes on large scales have the greatest magnitude. Further, as discussed in \S~\ref{sec:ilc_error} and shown in Fig.~\ref{fig:ffp8_n5diffmap_biased}, these errors are expected to increase in magnitude as a function of \(N\). This is because as \(N\) increases, each directional wavelet coefficient map (the space in which our ILC operates) contains fewer ``effective modes'' of the input data and so the error in our empirical covariance estimation is expected to increase. This error propagates to the final maps.

These map reconstruction errors due to the implementation of directionality on the very largest scales are accompanied by increasingly negative power spectrum residuals as \(N\) increases, in particular in the first multipole bin from \(\ell = 2\) to \(\ell = 11\). This is also indicative of the negative ILC bias due to empirical CMB cancellation, as discussed in \S~\ref{sec:ilc_error} and \citet{2009A&A...493..835D}. The results in Fig.~\ref{fig:ffp8_n5diffmap_biased} thus motivate the use of an axisymmetric scaling function, which ensures that no directionality is used for \(\ell < 32\) and so reduces the errors in CMB reconstruction. In principle these biases can be estimated and corrected through large suites of simulations, which is beyond the scope of this work.

\section{Application to \textit{Planck} data}
\label{sec:data}

\begin{figure*}
\begin{tabular}{cc} 
\includegraphics[width=\columnwidth]{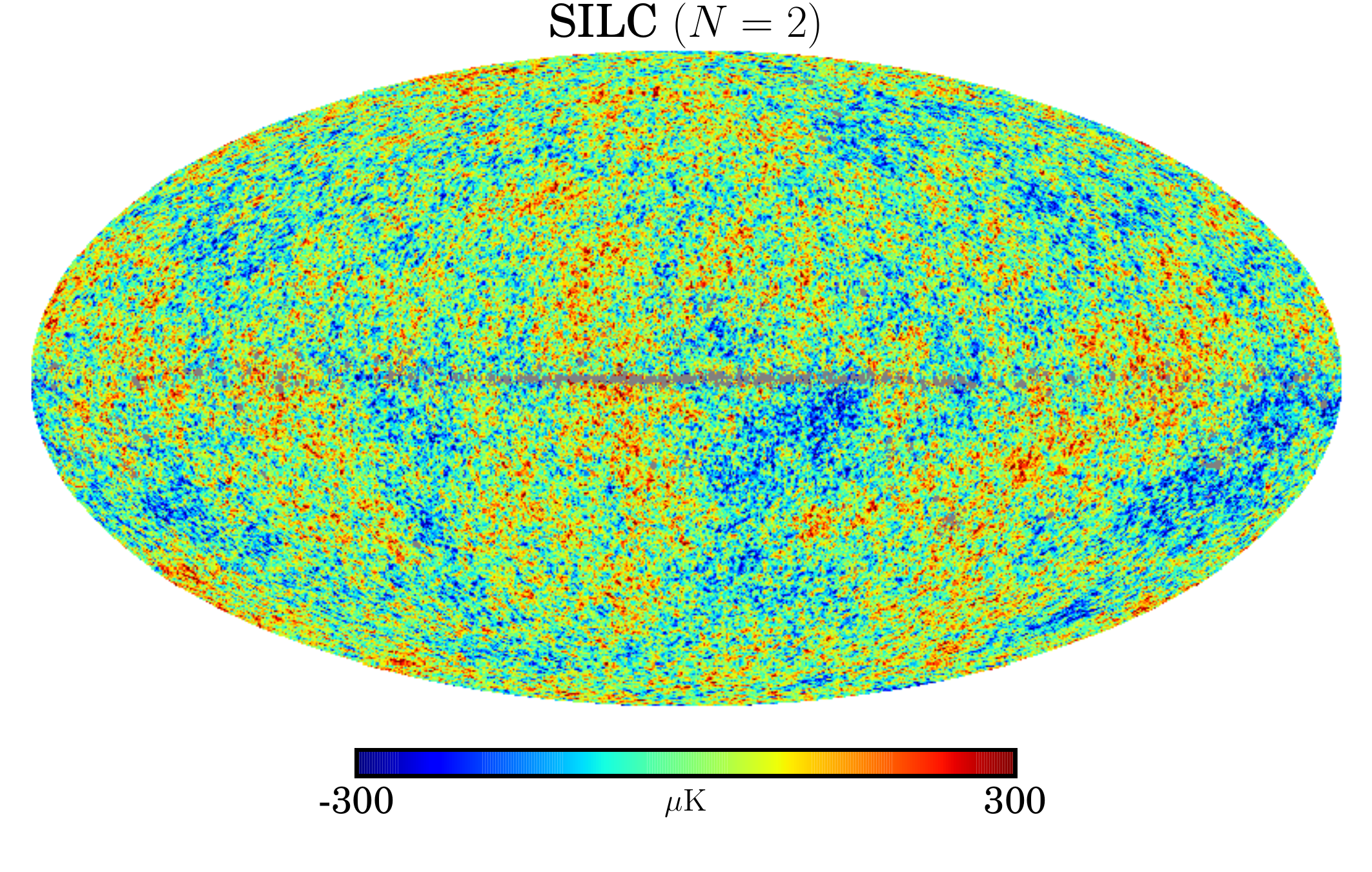} & \includegraphics[width=\columnwidth]{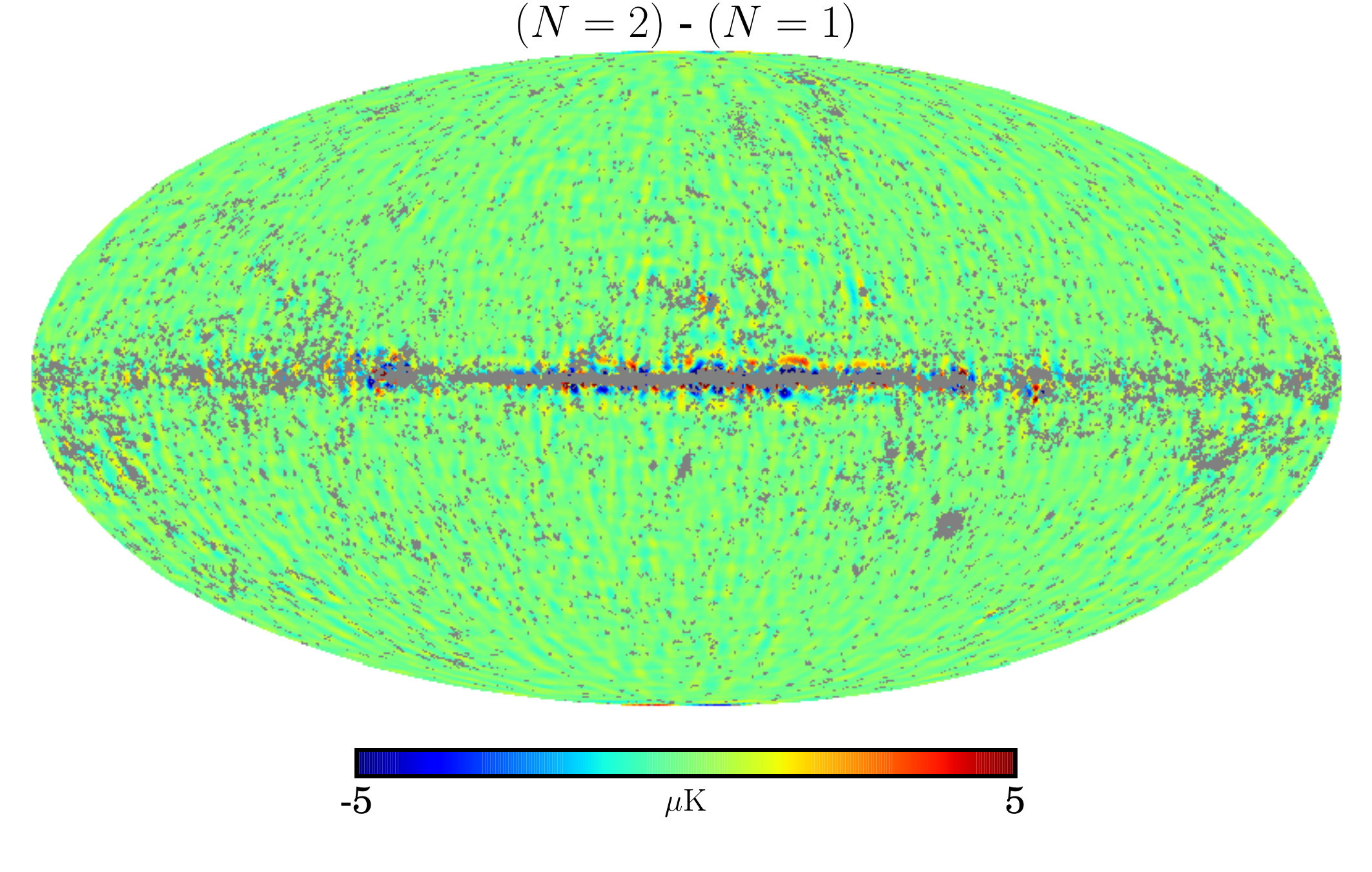} \\
\includegraphics[width=\columnwidth]{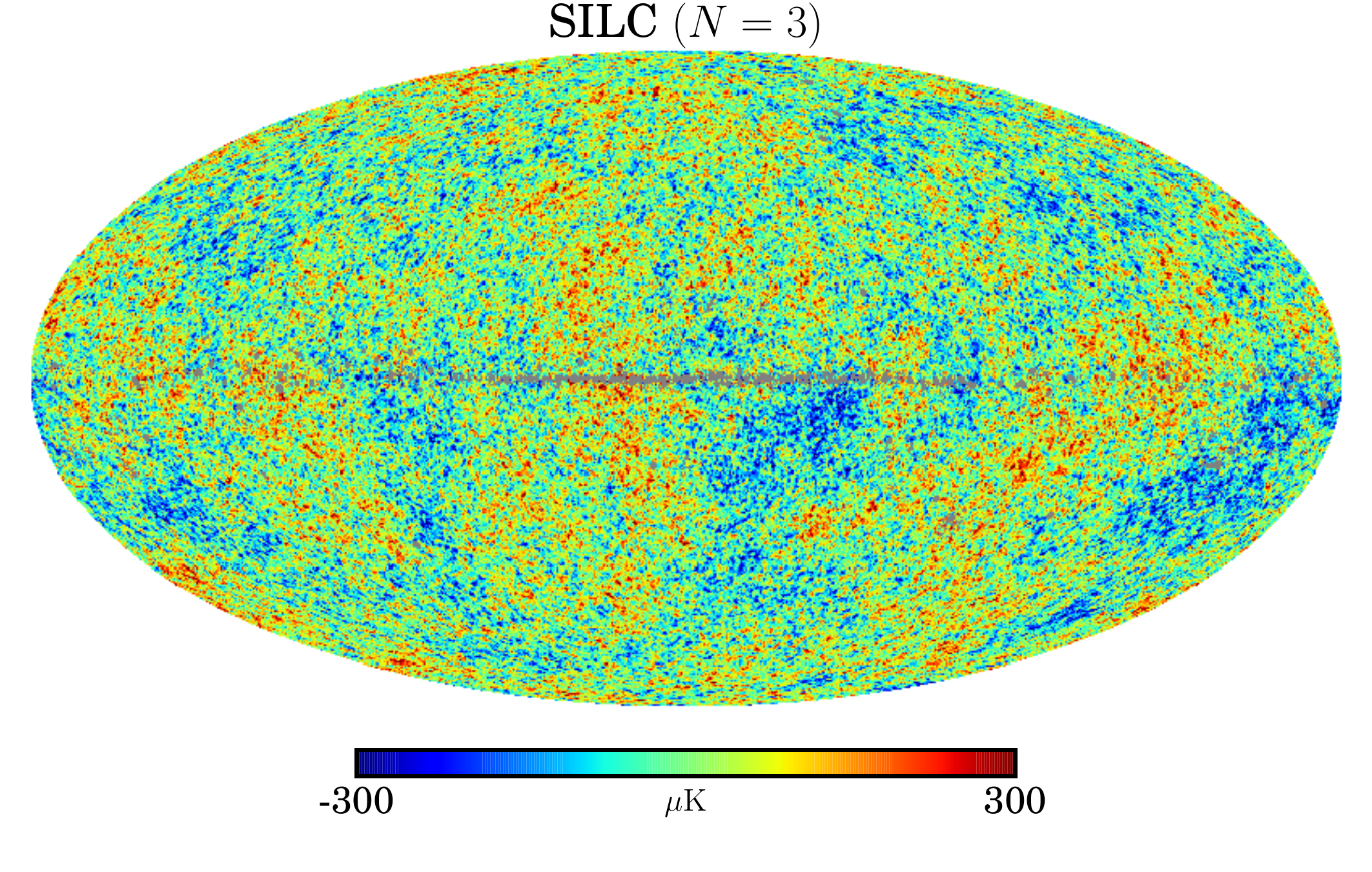} & \includegraphics[width=\columnwidth]{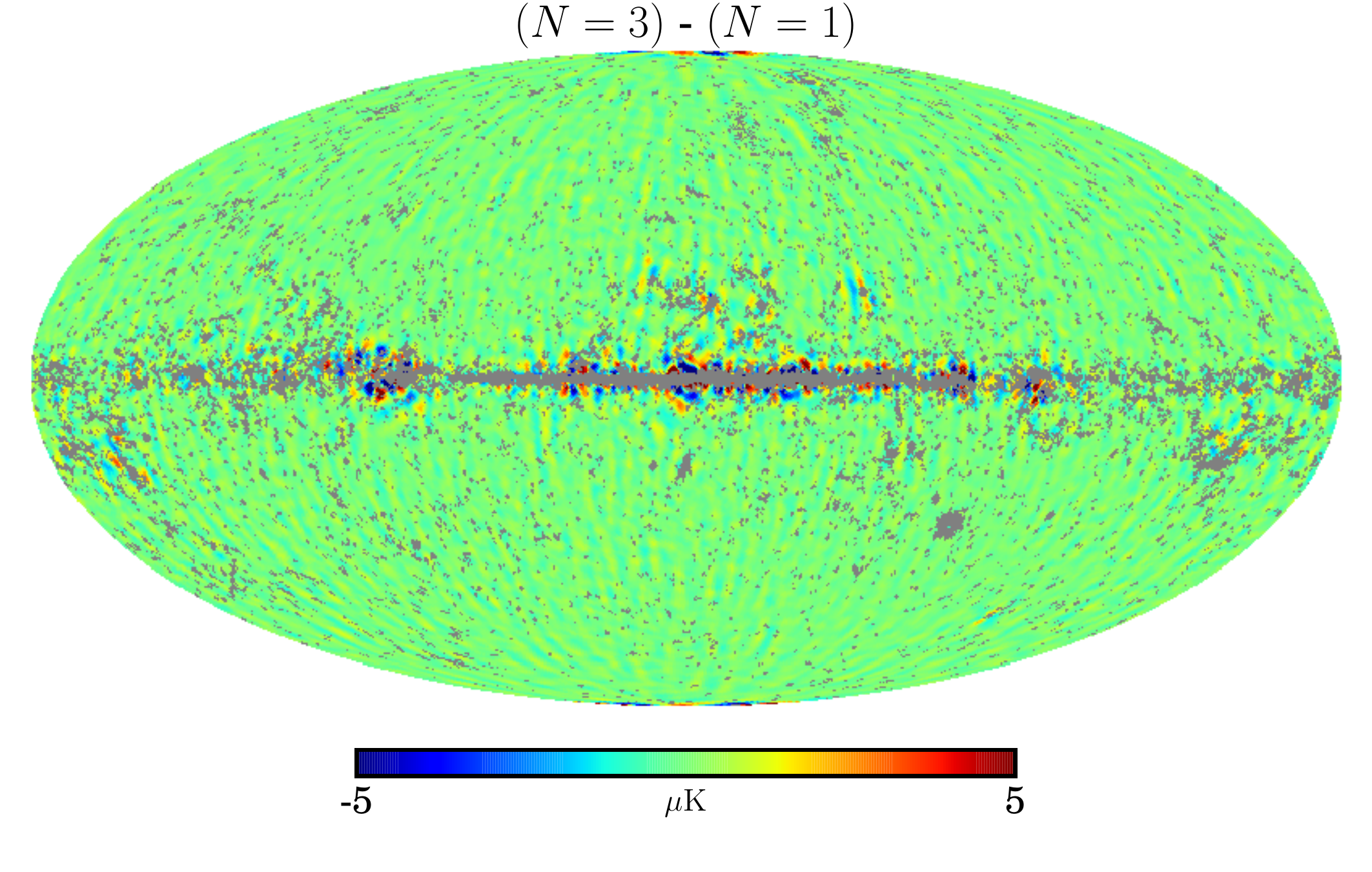} \\
\includegraphics[width=\columnwidth]{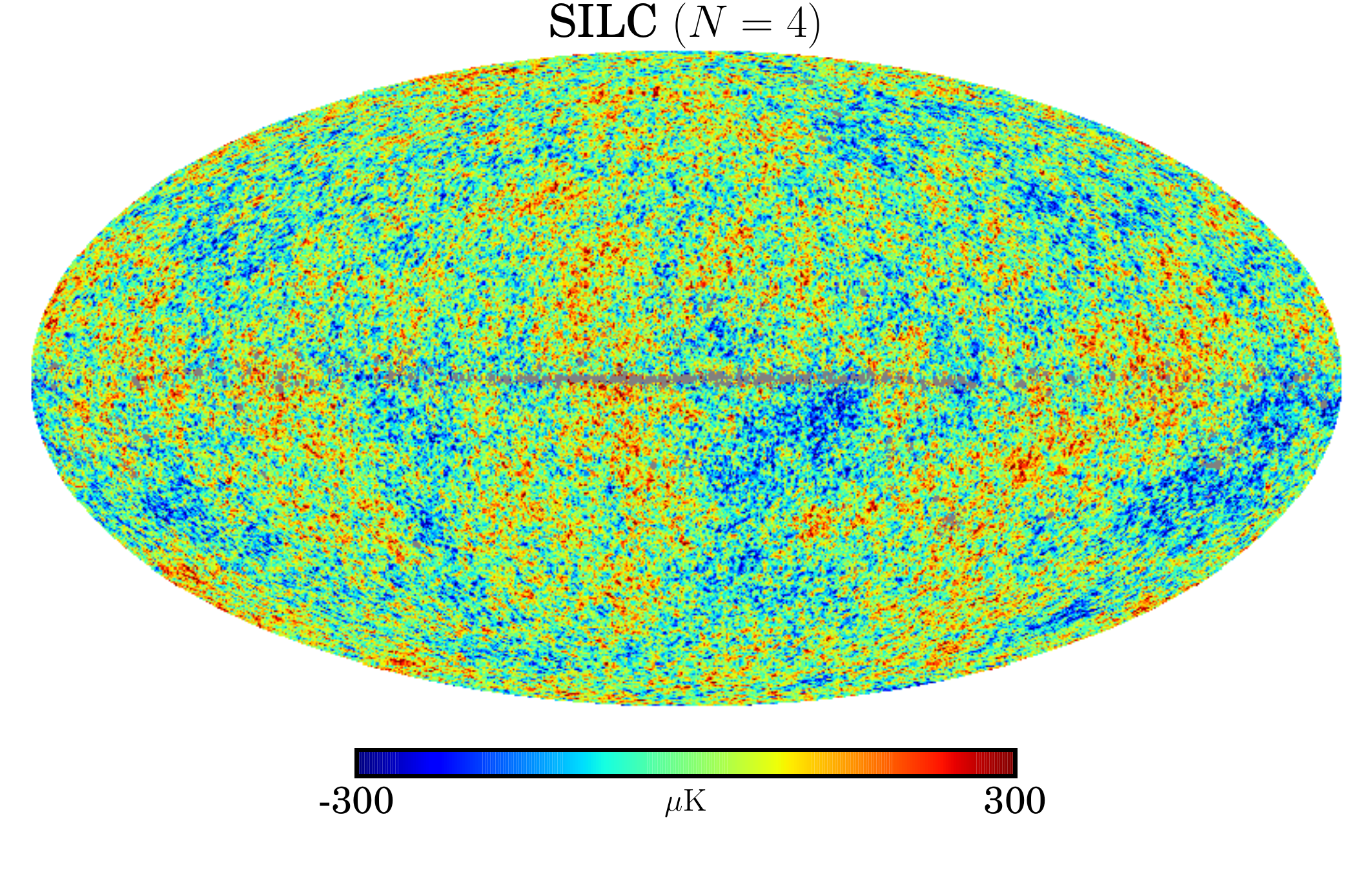} & \includegraphics[width=\columnwidth]{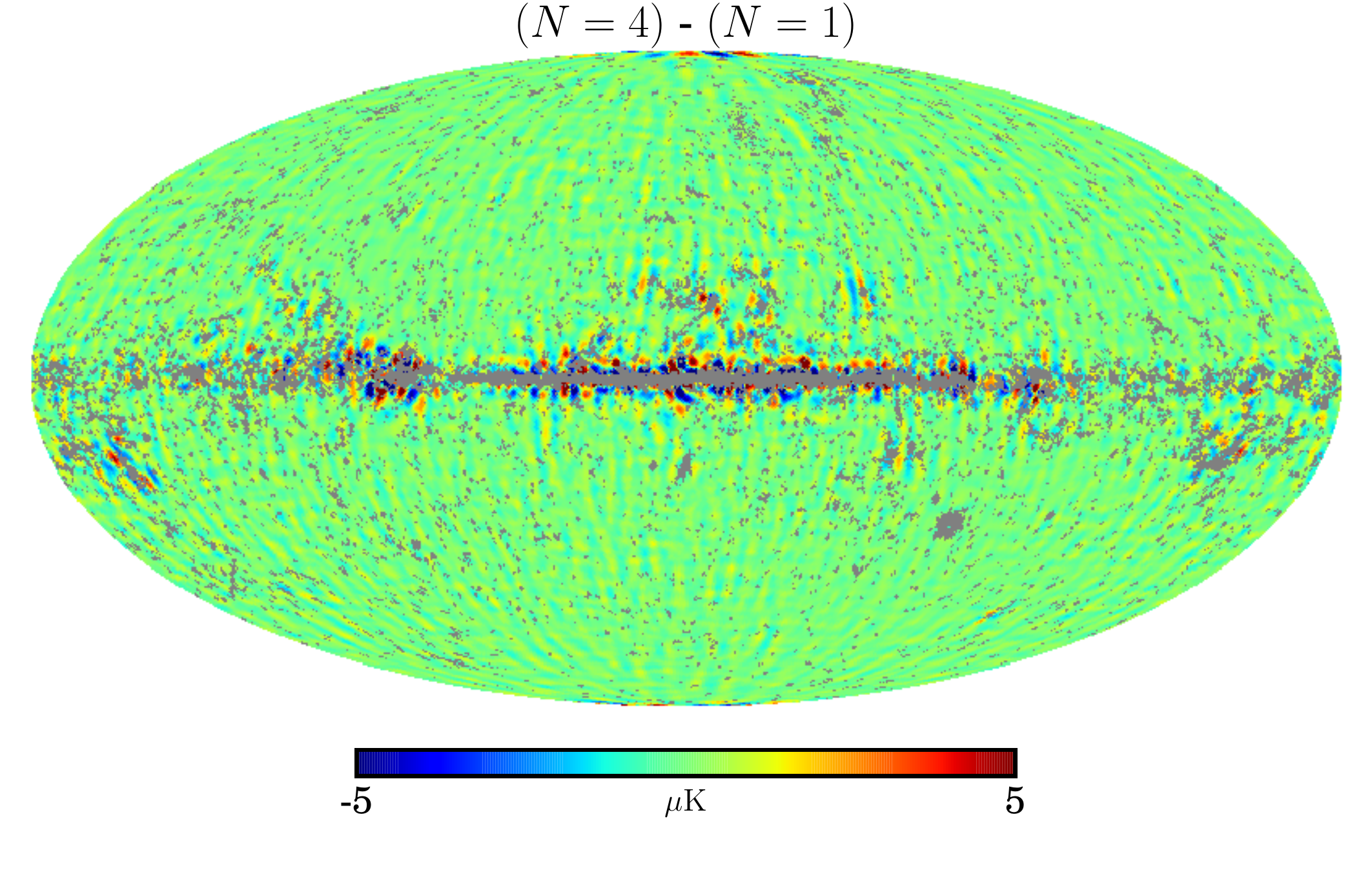} \\
\includegraphics[width=\columnwidth]{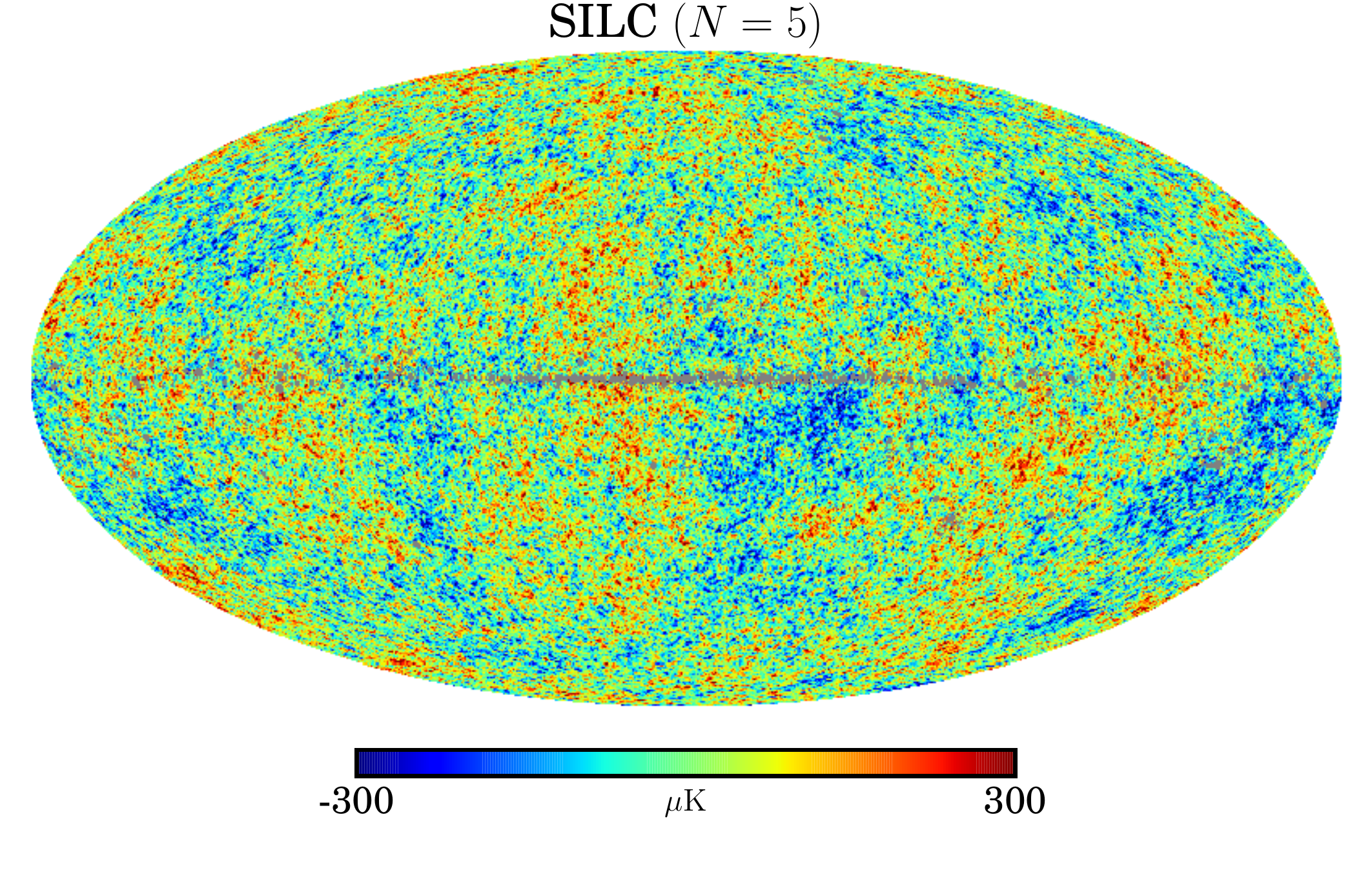} & \includegraphics[width=\columnwidth]{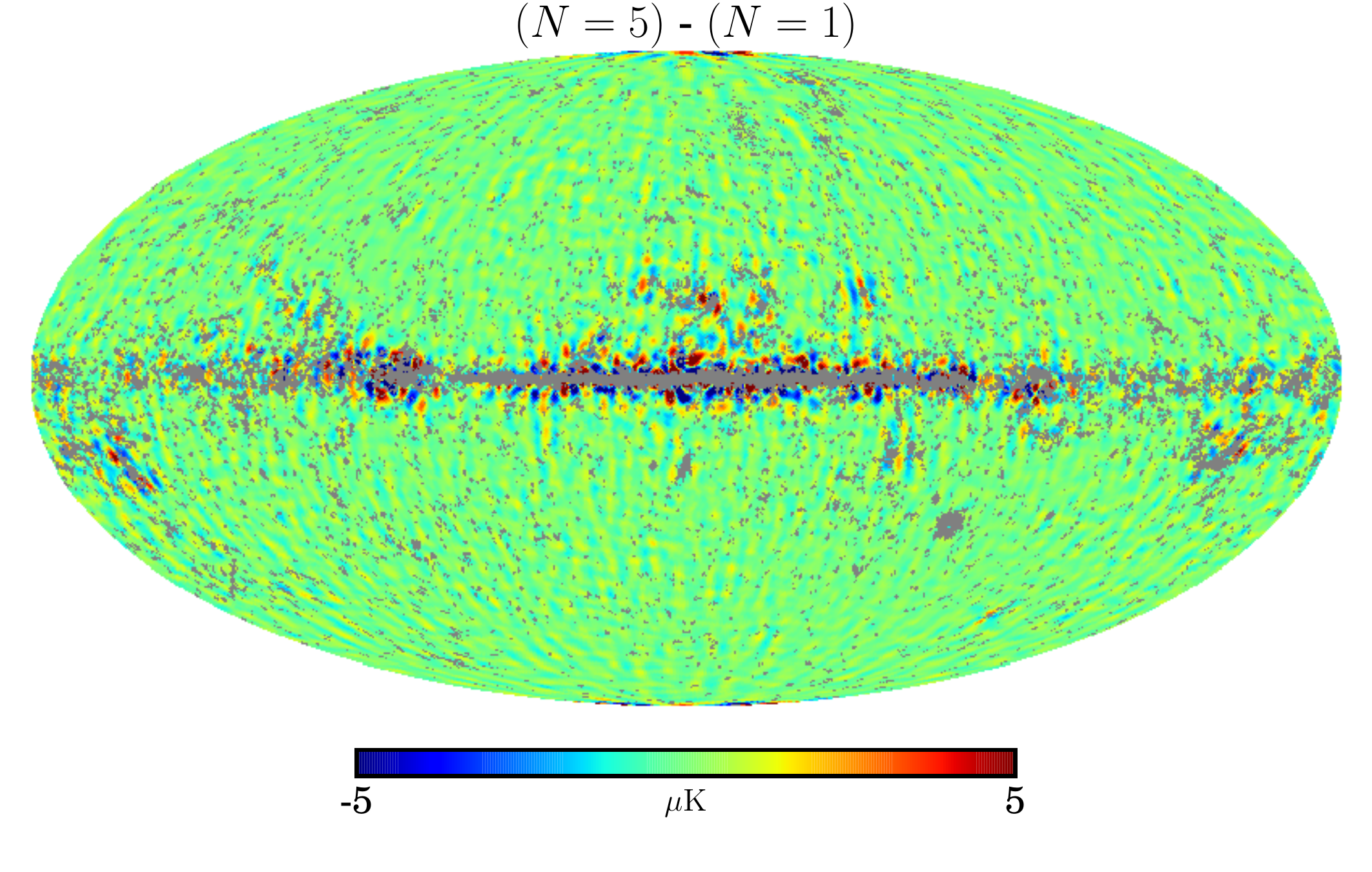}
\end{tabular}
\caption{Planck \textit{data.} \textit{Left}: CMB temperature anisotropies reconstructed using SILC with different values of \(N\) (FWHM = 5\arcmin, \(N_\mathrm{side} = 2048\)). \textit{Right}: differences between CMB temperature maps reconstructed using different values of \(N\) minus the axisymmetric limit \(N = 1\). The maps have been smoothed to \(\mathrm{FWHM} = 80\arcmin\) and downgraded to \(N_\mathrm{side} = 128\). \textit{In both columns}: the grey pixels are the point source mask (downgraded in resolution as appropriate). \textit{From top to bottom}: (a) \(N = 2\), (b) \(N = 3\), (c) \(N = 4\), (d) \(N = 5\).}
\label{fig:n5maps}
\end{figure*}

\renewcommand{\arraystretch}{0}
\begin{figure}
\begin{tabular}{c}
\includegraphics[width=\columnwidth]{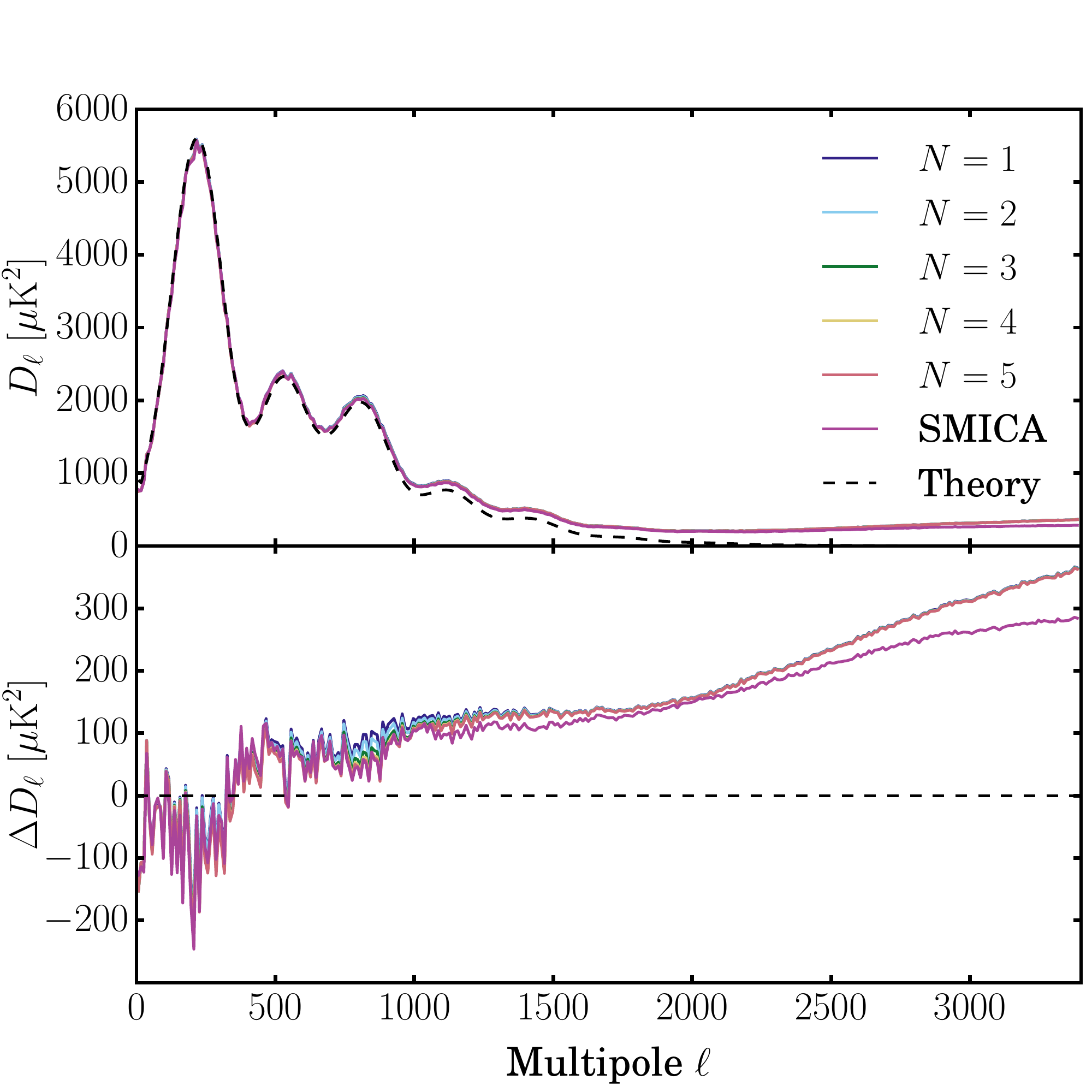} \\
\includegraphics[width=\columnwidth]{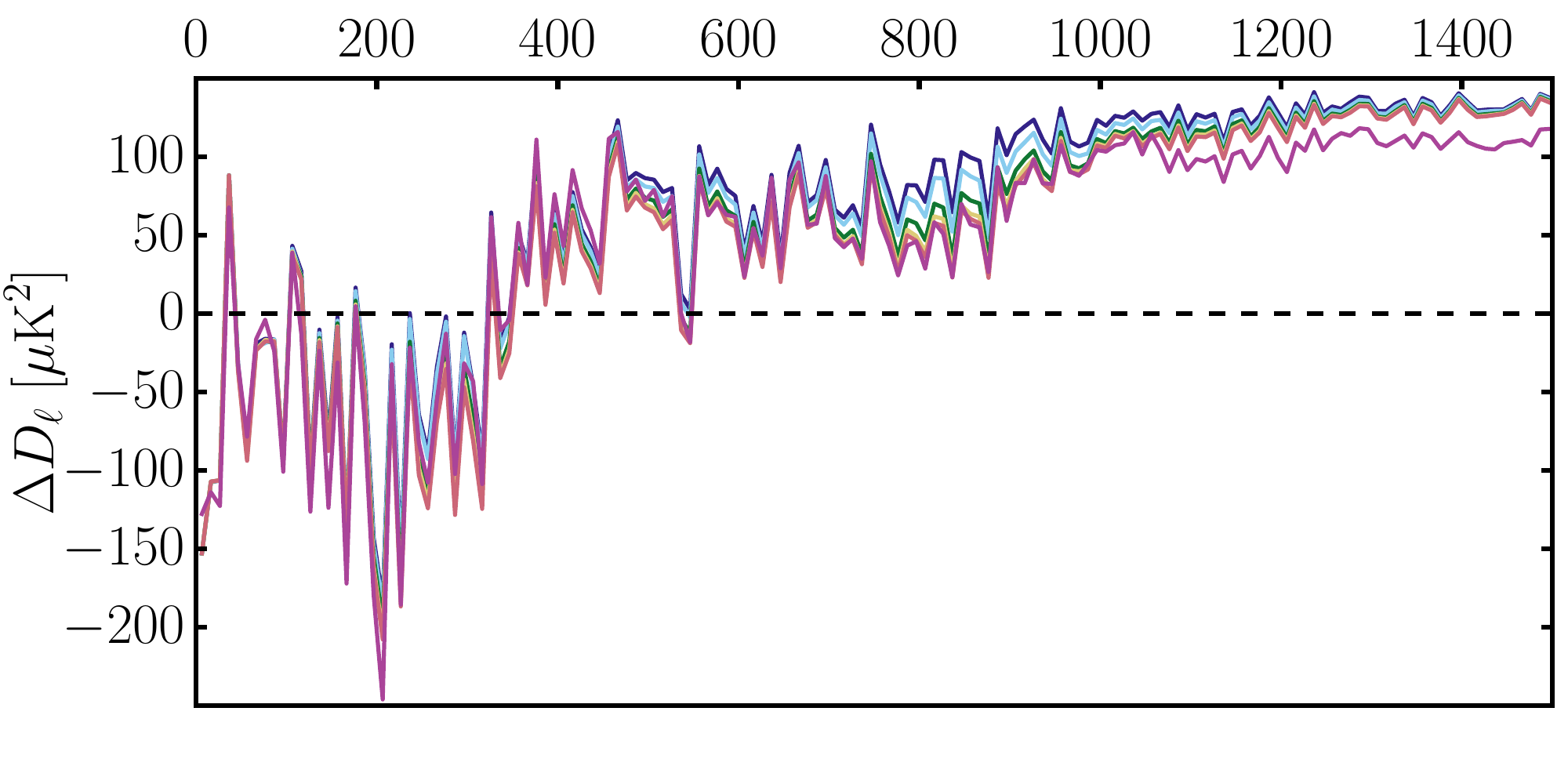}
\end{tabular}
\caption{Planck \textit{data.} \(TT\) angular power spectra comparing different values of \(N\) from 1 to 5 and SMICA. The top panel (a) shows point source masked spectra. The middle panel (b) shows residuals after subtracting the best-fit \(\Lambda\)CDM model from the {\Planck} 2015 likelihood. The bottom panel (c) shows the same residuals at low multipoles only (\(\ell < 1500\)).}
\label{fig:n5spec}
\end{figure}
\renewcommand{\arraystretch}{1}

We now study the application of SILC with increasing directionality to the full-mission {\Planck} sky maps.  The left column of Fig.~\ref{fig:n5maps} shows the full-resolution reconstructed CMB maps as calculated with different values of \(N\) from 2 to 5, which visually appear very similar. The right column of Fig.~\ref{fig:n5maps} shows the differences between the CMB reconstructed using \(N = [2,3,4,5]\) minus the axisymmetric limit (when \(N = 1\)), highlighting the differences at the lowest multipoles. The differences are of largest magnitude towards the Galactic plane where foreground emission is concentrated. This shows how the different wavelet kernels are localising the ILC weights differently in response to the directional structure of the foregrounds and CMB. The differences are small, reflecting the implementation of an axisymmetric scaling function, meaning that no directionality is applied at \(\ell < 32\). Figure \ref{fig:n5spec} compares point source masked \(TT\) angular power spectra of the CMB reconstructed using values of \(N\) from 1 to 5. The power spectrum residuals from a {\Planck} best-fit \(\Lambda\)CDM model remain small for most scales until the reconstructed spectra reach a characteristic noise spectrum for \(\ell \goa 1500\) where the different values of \(N\) converge. At these high multipoles, the ILC solution is dominated by residual instrumental noise. We see the biggest impact from directionality at intermediate multipoles (from \(\ell = 400\) to \(\ell = 1500\)). For comparison, we plot the SMICA power spectrum. In further support to the discussion in \S~\ref{sec:comp}, SILC matches the performance of SMICA. We note that, as with the simulations in \S~\ref{sec:simulations}, directionality changes the reconstructed CMB power spectrum most significantly at intermediate multipoles around \(\ell = 800\).

\section{Discussion}
\label{sec:discussion}

The comparisons in \S~\ref{sec:comp} demonstrate that SILC matches the performance of two previous methods, NILC and SMICA, in both maps and power spectra, with particular similarity between the axisymmetric limit of SILC and NILC, as expected. Both map residuals and power spectra in \S~\ref{sec:simulations} and \S~\ref{sec:data} show that switching on directionality changes CMB reconstruction most significantly at intermediate multipoles $\ell = 400$ -- $1500$. There appears to be little benefit in localising the ILC with directional wavelets at the very smallest scales, where the ILC result is noise-limited. We also adopt an axisymmetric scaling function on the very largest scales, meaning that there is no directionality at \(\ell < 32\). In Fig.~\ref{fig:ffp8_n5diffmap_biased}, we show the large CMB reconstruction errors arising from using directionality on the largest scales. This motivates the use of an axisymmetric scaling function, which significantly reduces the errors as seen in Fig.~\ref{fig:ffp8_n5diffmap}. In \S~\ref{sec:simulations}, we sketched out an argument that attributes these errors to empirical CMB cancellation (\S~\ref{sec:ilc_error}). However, the precise source and exact magnitude of any ILC errors are best estimated through suites of simulations.

We have presented this analysis by producing CMB maps (in \S~\ref{sec:simulations} and \S~\ref{sec:data}) each with a different single value of \(N\) at all wavelet scales. Our method can be simply extended to allow different values of \(N\) at each wavelet scale. In the same way that each wavelet scale has different harmonic band-limits, they can also have different azimuthal band-limits, optimised as identified above to reduce foreground and noise residuals. 

The negative ILC power spectrum biases discussed in \S~\ref{sec:ilc_error} must be quantified in parallel to this directionality optimisation if using the resulting map for power spectrum analyses.  It is possible to estimate variance biases in the data through suites of realistic simulations. However, we can also calculate this using the data itself and a fiducial CMB spectrum. In wavelet space, the variance estimator at each wavelet coefficient is \(\langle W^\mathrm{ILC} W^{\mathrm{ILC}\dagger}\rangle = \vec{\omega}^\dagger \langle \vec{W} \vec{W}^\dagger \rangle \vec{\omega} = \vec{\omega}^\dagger \mat{R} \vec{\omega} = (\vec{a}^\dagger \mat{R}^{-1} \vec{a})^{-1} = (\sum_{c,c'=1}^{N_c} (R^{-1})^{cc'})^{-1}\), where each equality follows by applying in turn Eqs.~\eqref{eq:ilc_sum}, \eqref{eq:covar_approx} and \eqref{eq:weights} (from \S~\ref{sec:ilc}) and then expanding the vector notation (the vectors span the space defined by the number of input channels; explicitly, we assume unit CMB calibration \(\vec{a} = (1,1,1,1,1,1,1,1,1)\)). In order to calculate the variance bias in the ILC, we can subtract the expected CMB variance \(\sum_{\ell m} C_\ell {| \Psi_{\ell m}^j |}^2\). If analysis was done in wavelet space, the above would define the variance bias. If only considering the diagonal terms in wavelet space, it is possible to straightforwardly transform this estimate to real space through an inverse wavelet transform as in Eq.~\eqref{eq:wav_syn}, substituting \(\Phi (\hat{\vec{n}})\) and \(\Psi^j (\hat{\vec{\rho}})\) respectively for \({| \Phi (\hat{\vec{n}}) |}^2\) and \({| \Psi^j (\hat{\vec{\rho}}) |}^2\). However, for a full treatment of the variance bias including off-diagonal terms, full wavelet space covariances need to be calculated. Although many off-diagonal terms would decay, this would still be computationally demanding and will be the focus of future research. However, we reiterate that if analysis is carried out in wavelet space, then variance biases can be straightforwardly calculated from the information already contained in the results.

\section{Conclusions}
\label{sec:concs}

We have presented SILC, a new form of internal linear combination that uses directional, scale-discretised wavelets to localise the ILC weighting according to the frequency, spatial, harmonic and, for the first time, morphological information in the CMB and its foregrounds. This is motivated by the anisotropic or filmentary morphology of both the CMB and astrophysical foregrounds in the microwave sky. We have tested SILC on 2015-release {\Planck} data and simulations, demonstrating comparable performance to two existing component separation algorithms, NILC and SMICA, and investigated how to optimise the use of morphological information through directionality. We have explored increasing the amount of directionality in the algorithm, showing that on the largest and the smallest scales, the axisymmetric limit of the ILC works well, while at intermediate multipoles (from $\ell = 400$ -- $1500$), increasing \(N\) (the number of directions per scale) leads to lower residuals. At high multipoles (\(\ell \goa 1500\)), the input data are already noise-limited, as is the ILC reconstruction, and directionality does not reduce the reconstruction error, as instrumental noise has no directional structure. We adopt an axisymmetric scaling function to analyse the largest scales, meaning that we use no directionality for \(\ell < 32\). This is motivated by the observation that increasing directionality on large scales gives increased reconstruction errors over the axisymmetric limit. We argue that these errors are due to empirical CMB cancellation in the ILC calculation, though the exact source must be estimated through large suites of realistic simulations. Allowing \(N\) to vary with wavelet scale is analogous to the choice of different harmonic band-limits at different scales.

We conclude that the introduction of directional wavelets allows greater flexibility in the ILC to make use of morphological information at targeted scales. Our multiprocessing implementation takes advantage of the wavelet scales to allow large-scale results to be analysed while small scales are still being processed. Moreover, our wavelet transforms are quick and exact, using MW sampling and fast Fourier transforms (see \S~\ref{sec:dir_wav}). As discussed in \S~\ref{sec:ilc_error}, the ILC is prone to several sources of error and variance bias, including empirical CMB cancellation. This bias can be estimated through suites of Monte Carlo simulations, but we have also outlined (in \S~\ref{sec:discussion}) the ability to estimate biases directly from the data, most straightforwardly in wavelet space. We make our map products available at \url{http://www.silc-cmb.org}\footnote{The DOI for our data release is 10.5281/zenodo.44373.}.

This work on scalar signals (\ie the temperature \(I\) component of the CMB) can be extended to spin signals (\ie the polarisation \(Q\) and \(U\) components of the CMB, or, equivalently, the \(E\) and \(B\) modes), by using spin wavelets \citep{mcewen:s2let_spin, 2014IAUS..306...64M,2015arXiv150203120L}. These are an extension of directional, scale-discretised wavelets to represent spin signals, such as CMB polarisation, a spin $\pm 2$ signal. We expect that the directionality will be particularly suited to the anisotropic, filamentary nature of polarised foregrounds when observed on the sky, and in future work will present the application of SILC to CMB polarisation data.

\section*{Acknowledgements}
\label{sec:ack}

KKR thanks Franz Elsner for valuable discussions. KKR was supported by the Science and Technology Facilities Council. HVP and BL were partially supported by the European Research Council under the European Community's Seventh Framework Programme (FP7/2007-2013) / ERC grant agreement number 306478-CosmicDawn. JDM was partially supported by the Engineering and Physical Sciences Research Council (grant number EP/M011852/1). AP was supported by the Royal Society. Based on observations obtained with {\Planck} (\url{http://www.esa.int/planck}), an ESA science mission with instruments and contributions directly funded by ESA Member States, NASA and Canada.

\bibliographystyle{mymnras_eprint}
\bibliography{bib_journal_names_short,s2let_ilc_mnras}

\end{document}